\documentclass[10pt]{iopart}

\usepackage{iopams}  

\expandafter\let\csname equation*\endcsname\relax
\expandafter\let\csname endequation*\endcsname\relax

\usepackage{
,amsfonts,
amssymb,
amsthm,
amsmath} 

\usepackage{rawfonts}

\let \latexput\put
\input prepictex
\let \put\pictexput
\input pictex
\input postpictex

\let \put\latexput

\pdfoutput=1
\usepackage{etex}
\usepackage{cite}
\usepackage{latexsym}
\usepackage{graphicx}
\usepackage[usenames,dvipsnames,svgnames]{xcolor}
\usepackage{bbm}
\usepackage{latexsym}
\usepackage{multirow}
\usepackage{rotating}
\usepackage{lscape}
\usepackage{graphicx} 
\usepackage[format=plain,labelfont=it,textfont=it]{caption}
\usepackage[format=plain,labelfont=it,textfont=it]{subcaption}
\usepackage{xcolor}
\usepackage{fancybox}
\usepackage{ifmtarg}
\usepackage{fancyhdr}
\usepackage{wrapfig}
\usepackage{hyperref}
\usepackage{tikz}
\usepackage{bm}
\usepackage[flushleft]{threeparttable}
\usepackage{cases}

\def\2;{\;\;}

\def\eps{\epsilon}
\def\P#1#2{{P^{(#1)}_{#2}}}
\def\Pr#1{{\mathbb{P}^{(#1)}}}

\def\IntZ{{\mathbb Z}}

\def\RealN{{\mathbb R}}

\def\mathL{{\mathbb L}}

\def\mathP{{\mathbb P}}

\def\o#1{\overline{#1}}

\def\Ref#1{(\ref{#1})}

\def\binom#1#2{{{#1}\choose{#2}}}
\def\Bi#1#2{{\binom{#1}{#2}}}

\def\Sfrac#1#2{\hbox{\large $\frac{#1}{#2}$}}
\def\sfrac#1#2{\hbox{\nor $\frac{#1}{#2}$}}

\def\LB{\left(}         \def\RB{\right)}

\def\lfl{\!\left\lfloor} \def\rfl{\right\rfloor\!}

\def\nor{\normalsize}

\def\thin{ {\hspace{0.75pt}} }

\def\minus{{\hspace{0.85pt}{-}\hspace{0.85pt}}}

\def\tiny#1{\scalebox{0.75}{#1}}

\def\dps{\displaystyle}
\newcommand*{\erf}[0]{\hbox{erf}}
\newcommand*{\erfc}[0]{\hbox{erfc}}

\def\Pr{\mathbb{P}_r}

\begin{document}

\title[Trajectories of square lattice staircase polygons]{
Trajectories of square lattice staircase polygons}

\author{EJ Janse van Rensburg}
\address{\sf Department of Mathematics and Statistics, 
York University, Toronto, Ontario M3J~1P3, Canada\\}
\ead{\href{mailto:rensburg@yorku.ca}{rensburg@yorku.ca}}
\vspace{10pt}
\begin{indented}
\item[]\today
\end{indented}

\begin{abstract}
The distribution of monomers in a coating of grafted and adsorbing polymers
is modelled using a grafted staircase polygon in the square lattice.  The
adsorbing staircase polygon consists of a bottom and a top lattice path
(branches) and the asymptotic probability density that vertices in the
lattice are occupied by these paths is determined.  This is a model consisting
of two linear polymers grafted to a hard wall in a coating. The probability 
that either the bottom, or the top, path of a staircase polygon passes 
through a lattice site with coordinates $(\lfl \eps n \rfl,\lfl \delta \sqrt{n}\rfl)$, 
for $0 < \eps < 1$ and $\delta\geq 0$, is determined asymptotically 
as $n\to\infty$.  This gives the probability density of vertices in the
staircase polygon in the scaling limit (as $n\to\infty$): 
\[ \Pr^{f,(both)} (\eps,\delta) = \frac{\delta^2(15\,\eps^2(1\minus \eps)^2
- 12\, \delta^2 \eps(1\minus \eps) + 4\, \delta^4 )
}{3\sqrt{\pi\,\eps^7(1\minus\eps)^7}}
\, e^{-\delta^2/\eps(1\minus \eps)} . \]
The densities of other cases, grafted and adsorbed, and at the adsorption
critical point (or special point), are also determined.  In these cases 
the most likely and mean trajectories of the paths are determined.
The results show that low density regions in the distribution induces
entropic forces on test particles which confine them next to the hard wall,
or between branches of the staircase polygon.  These results give qualitative
mechanisms for the stabilisation of a drug particle confined to a polymer
coating in a drug delivery system such as a drug-eluding stent covered 
by a grafted polymer.  
\end{abstract}

%
\vspace{2pc}


%
%

\section{Introduction}
\label{s1}

\begin{figure}[h!]
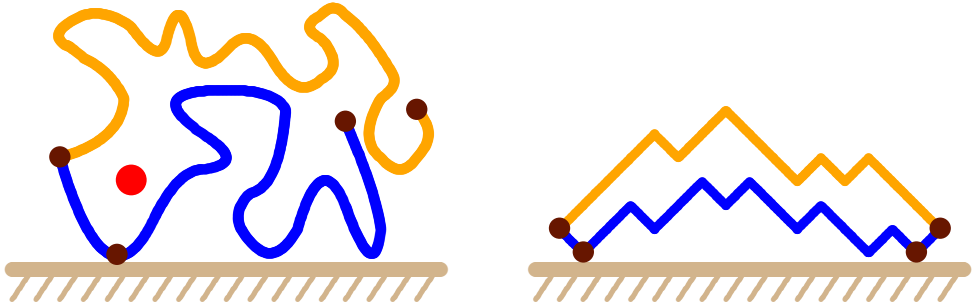

\beginpicture
\setcoordinatesystem units <0.9pt,0.9pt>
\setplotarea x from -30 to 250, y from -20 to 120

\color{Tan}
\setplotsymbol ({\Large$\bullet$})
\plot -20 -7 160 -7 /
\setplotsymbol ({\LARGE$\cdot$})
\plot -20 -20 -10 -5 / \plot -10 -20 0 -5 / \plot 0 -20 10 -5 / \plot 10 -20 20 -5 /
\plot 20 -20 30 -5 / \plot 30 -20 40 -5 / \plot 40 -20 50 -5 / \plot 50 -20 60 -5 /
\plot 60 -20 70 -5 / \plot 70 -20 80 -5 / \plot 80 -20 90 -5 / \plot 90 -20 100 -5 /
\plot 100 -20 110 -5 / \plot 110 -20 120 -5 / \plot 120 -20 130 -5 / \plot 130 -20 140 -5 /
\plot 140 -20 150 -5 / \plot 150 -20 160 -5 /

\color{Blue}
\setplotsymbol ({$\bullet$})
\setquadratic
\plot 0 40 20 0 40 15  50 30 60 35  70 40 60 50 50 65 80 68
90 65 95 60 90 35 80 25  75 15 80 5 90 0 100 10
110 30 120 20 130 0 135 10  130 30 120 55  /

\color{Orange}
\plot 0 40 20 50 27 65 20 75  10 82  5 86  3 87  0 93  10 100   20 100  30 95 
40 85  45 90  50 100   55 90  60 80  70 85  80 90  90 80  100 70  110 73
115 80 110 90  105 95  110 100 115 103 120 100  130 90 140 75  140 70
135 65 130 50 135 40 143 35 150 40 155 50 150 60 /

\color{Red}
\put {\scalebox{3.0}{$\bullet$}} at 30 30

\setlinear
\color{Sepia}
\multiput {\huge$\bullet$} at 24 -1 0 40 120 55 150 60 /

\setcoordinatesystem units <0.9pt,0.9pt> point at -220 0
\setplotarea x from -15 to 250, y from -20 to 80

\color{Tan}
\setplotsymbol ({\Large$\bullet$})
\plot -20 -7 160 -7 /
\setplotsymbol ({\LARGE$\cdot$})
\plot -20 -20 -10 -5 / \plot -10 -20 0 -5 / \plot 0 -20 10 -5 / \plot 10 -20 20 -5 /
\plot 20 -20 30 -5 / \plot 30 -20 40 -5 / \plot 40 -20 50 -5 / \plot 50 -20 60 -5 /
\plot 60 -20 70 -5 / \plot 70 -20 80 -5 / \plot 80 -20 90 -5 / \plot 90 -20 100 -5 /
\plot 100 -20 110 -5 / \plot 110 -20 120 -5 / \plot 120 -20 130 -5 / \plot 130 -20 140 -5 /
\plot 140 -20 150 -5 / \plot 150 -20 160 -5 /

\color{Blue}
\setplotsymbol ({\footnotesize$\bullet$})
\plot -10 10 0 0  10 10  20 20  30 10 40 20 50 30 60 20 70 30 80 20
90 10 100 20 110 10 120 0 130 10 140 0 150 10 /
\color{Orange}
\plot -10 10 0 20 10 30  20 40  30 50  40 40 50 50 60 60  70 50 80 40 
90 30  100 40 110 30 120 40 130 30 140 20 150 10  /
\color{Sepia}
\multiput {\huge$\bullet$} at 0 0 140 0 -10 10 150 10 /

\color{Black}
\normalcolor
\endpicture
\caption{(Left) A schematic diagram of a linear polymers in a coating
confining a particle. In this diagram a 2-block copolymer is shown, with one 
of the blocks grafted to the wall.  Steric repulsions between the two blocks 
and the hard wall create low density regions in the coating.  One block may 
adsorb on the surface.  (Right) A staircase polygon model of a grafted 
2-block ring copolymer.  The bottom block is grafted at both its 
endpoints to the wall (and may adsorb on the wall), while the top block 
is screened from the surface by the bottom block. Each of the blocks 
in the staircase polygon is a directed path giving North-East or South-East 
in the square lattice.}
\label{F1}
\end{figure}

Polymer coatings on surfaces or on suspended colloid particles have 
applications in medical and industrial systems.  For example, there are
important uses in targeted drug delivery systems \cite{S06,JM12,LKSP10},
including nanoparticle-polymer systems \cite{TOS01,SYS15}.  Polymer 
coatings on drug-eluding stents are now widely used to deliver a drug 
over a period time \cite{FHS01} at the location of the stent.  In other
examples, the steric stabilisation of colloid dispersions by coating particles 
with a polymer are important in a variety of applications \cite{FS93,N83}.
In this case the colloid particles are coated with a polymer and, if they 
approach each other, experience repulsive forces due to the reduction 
in configurational entropy of the polymers between them.  In nano-particle 
systems and drug eluding stents a particle is suspended in a polymer coating
and it may diffuse from the coating over time, or be released as the coating 
degrades.  Copolymers are frequently used in these systems, and
in figure \ref{F1} (left panel) a schematic diagram of a $2$-block copolymer
grafted to a surface is shown.  The copolymer coils near the surface
to form a coating into which a drug absorbs to be released later at a 
target site. 

In these applications the properties of the polymeric coating plays an important
role.  A hydrophylic coating has polymer coils that form a thicker and 
less dense layer, while a hydrophobic coating results in a thinner and 
denser layer.  Both thinner and thicker coatings have their applications, 
and the density distribution of the layer (for example, less dense towards 
the wall, more dense away from it) are relevent.

In reference \cite{JvR23} a directed path model of a grafted linear polymer 
in a polymer coating was examined.  It was shown there that the entropic
repulsion between the hard wall and the directed path creates a low density
region next to the wall, providing a metastable location for absorbed particles 
next to the wall.  However, this region disappears when the path adsorbs 
on the wall, indicating that a grafted polymer (rather than an adsorbed 
polymer) would be more efficient in retaining particles in the coating.

In this paper a more elaborate directed path model of polymers in a polymer 
coating is examined.  In figure \ref{F1} (left panel) a schematic drawing
of a 2-block copolymer, with one block grafted to a hard wall, is shown.
The two blocks avoid each other and the wall, and if they approach each
other of the wall, then a reduction in configurational entropy induces 
mutually repulsive forces between them.   If the polymer is 
a 2-block ring copolymer, then it can be modelled by a \textit{grafted 
staircase polygon} shown in the right panel of figure \ref{F1}.  Two directed 
lattice paths, starting in the same vertex, giving North-East (NE) and South-East 
(SE) steps and avoiding each other, end in the same vertex to form a lattice
polygon.  The bottom directed path is one block, and it is grafted to the
wall at both endpoints, and may also adsorb on the wall if there is a strong
attractive force between the wall and the vertices in the path.  The top
path is the second block, and it is screened from the wall by the bottom path.
More generally, this is also a two dimensional model of two interacting 
linear polymers grafted to the wall, with one adsorbing onto the wall.

There is an extensive literature in the physical sciences on directed path
models of adsorbing and grafted polymers \cite{CP88,PFF88,F90,F91,W98}.  
These models are exactly solvable, and have proven particularly usefull
as a tool to quantify the entropy of a directed linear polymer and exposing
the general scaling and phase behaviour in linear polymers.   The simplest 
models are Dyck paths
\cite{BY95,BEO98,BORW05,RJvR04,JvR05,JvR10A,IJvR12,JvRP13} and these
are related to staircase polygons \cite{BG90,LZ91,JvRP13}.

\begin{figure}[t]
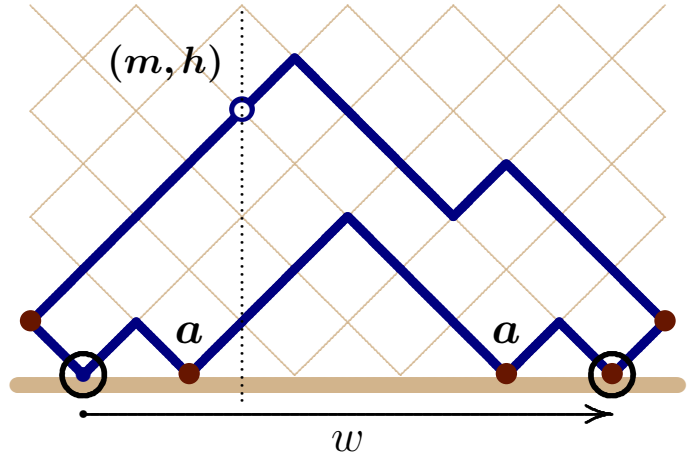

\centering
\beginpicture
\setcoordinatesystem units <1pt,1pt>
\setplotarea x from -90 to 220, y from -15 to 140

\color{Tan}
\plot 0 40 100 140 / \plot 0 80 60 140 / \plot 0 120 20 140 /
\plot 0 0 140 140 / \plot 40 0 180 140 / \plot 80 0 220 140 / 
\plot 120 0 220 100 / \plot 160 0 220 60 / \plot 200 0 220 20 / 

\plot 40 0 0 40 / \plot 80 0 0 80 / \plot 120 0 0 120 / \plot 160 0 20 140 /
\plot 200 0 60 140 / \plot 220 20 100 140 / \plot 220 60 140 140 /
\plot 220 100 180 140 /

\plot 0 0 -20 20 0 40 -20 60 0 80 -20 100 0 120 -20 140 /

\color{NavyBlue}
\setplotsymbol ({\footnotesize$\bullet$})

\plot -20 20 0 40 20 60 40 80 60 100 80 120 100 100 120 80 140 60 160 80 180 60 200 40 220 20 /
\plot -20 20 0 0 20 20 40 0 60 20 80 40 100 60 120 40 140 20 160 0 180 20 200 0 220 20 /

\setplotsymbol ({\scalebox{1.5}{$\bullet$}})
\color{Tan}
\plot -25 -4 225 -4 /

\color{Sepia}
\multiput {\huge$\bullet$} at -20 20 40 0 160 0 200 0 220 20  / 
\color{NavyBlue}
\multiput {\Large$\bullet$} at 0 0 / 
\multiput {\scalebox{2.5}{$\bullet$}} at 60 100 / 
\color{White}
\multiput {\scalebox{1.5}{$\bullet$}} at 60 100 / 

\color{Black}
\multiput {\LARGE$\bm{a}$} at 40 16 160 16  /

\put{\Large$\bm{(m,h)}$} at 31 118

\setplotsymbol ({\scalebox{0.5}{$\bullet$}})
\circulararc 360 degrees from 0 8 center at 0 0 
\circulararc 360 degrees from 200 8 center at 200 0 

\normalcolor\color{black}

\setplotsymbol ({$\cdot$})
\arrow <10pt> [.2,.67] from 0 -15 to 200 -15
\multiput {\scalebox{0.75}{$\bullet$}} at 0 -15 / 
\put {\LARGE$w$} at 100 -25

\setdots <3pt>
\plot 60 -10 60 140 /

\endpicture
\caption{A staircase polygon grafted at the origin and at the end of its bottom
path.  The width $w$ of the polygon is equal to the length of its bottom
path minus $2$.  Each return of the bottom path to the hard wall, except the origin
and the last grafted vertex, is weighted by $a$.  The top path passes 
through the vertex with coordinates $(m,h)$ and by calculating the probability
of events like this, the probability density of the polygon is obtained in the
limit as its length increases to infinity.  The width of the polygon is the 
distance from the origin to the penultimate vertex in the bottom path, 
or the number of steps between the origin and the penultimate vertex in 
the bottom path.}
\label{F2}
\end{figure}

\subsection{Adsorbing staircase polygon models}

The \textit{length} of a directed lattice path giving NE and SE steps in the 
square lattice $\mathL$ is its number of steps.  Since the path always steps 
one unit per step in the East direction, the length of the path is also 
its \textit{width}.  The length of a staircase polygon is the sum of the lengths 
of its bottom and top paths (see figure \ref{F2}), or twice the length of either 
its bottom or top path.  A staircase polygon is \textit{grafted at both its
endpoints} if the second and penultimate vertices in the bottom path are fixed 
in the hard wall.   The width of a grafted staircase polygon is equal to
the length of its bottom walk \textit{minus $2$}, as illustrated in figure
\ref{F2}.   If \textit{returns} of the bottom path to the wall (except 
for the grafted vertices) are weighted by a Boltzmann factor 
$a=e^{-\mathcal{E}/kT}$, then the staircase polygon is \textit{adsorbing} 
and $\mathcal{E}$ is a binding energy of a vertex to the hard wall, 
$T$ is the absolute temperature and $k$ is Boltzmann's constant.

\begin{figure}[t]
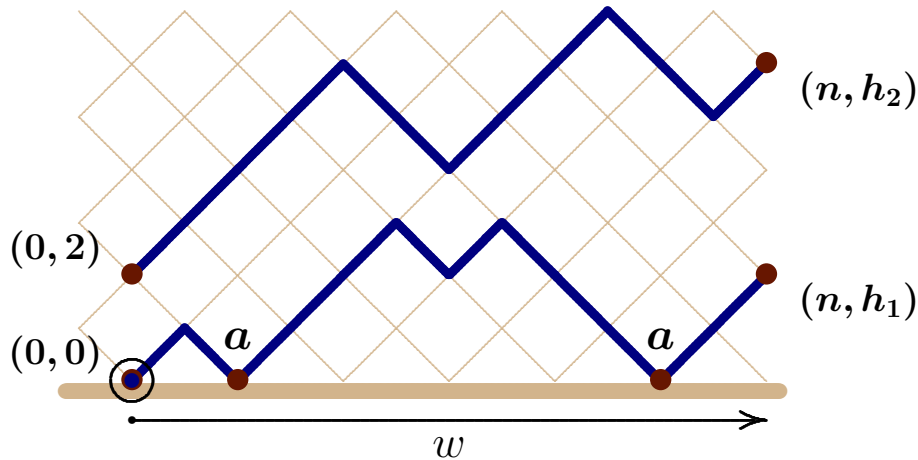

\centering
\beginpicture
\setcoordinatesystem units <1pt,1pt>
\setplotarea x from -80 to 220, y from -15 to 140

\color{Tan}
\plot 0 40 100 140 / \plot 0 80 60 140 / \plot 0 120 20 140 /
\plot 0 0 140 140 / \plot 40 0 180 140 / \plot 80 0 220 140 / 
\plot 120 0 220 100 / \plot 160 0 220 60 / \plot 200 0 220 20 / 

\plot 40 0 0 40 / \plot 80 0 0 80 / \plot 120 0 0 120 / \plot 160 0 20 140 /
\plot 200 0 60 140 / \plot 220 20 100 140 / \plot 220 60 140 140 /
\plot 220 100 180 140 /

\plot 0 0 -20 20 0 40 -20 60 0 80 -20 100 0 120 -20 140 /
\plot 240 0 220 20 240 40 220 60 240 80 220 100 240 120 220 140 /

\color{NavyBlue}
\setplotsymbol ({\footnotesize$\bullet$})

\plot 0 40 20 60 40 80 60 100 80 120 100 100 120 80 140 100 160 120 180 140 200 120
220 100 240 120 /
\plot 0 0 20 20 40 0 60 20 80 40 100 60 120 40 140 60 160 40 180 20 200 0 220 20 240 40 /

\setplotsymbol ({\scalebox{1.5}{$\bullet$}})
\color{Tan}
\plot -25 -4 245 -4 /

\color{Sepia}
\multiput {\huge$\bullet$} at 0 0 40 0 200 0 240 40 240 120 0 40   / 
\color{NavyBlue}
\multiput {\Large$\bullet$} at 0 0 / 
\color{White}

\color{Black}
\multiput {\LARGE$\bm{a}$} at 40 16 200 16 /
\put {\Large$\bm{(n,h_1)}$} at 275 30 
\put {\Large$\bm{(n,h_2)}$} at 275 110 
\put {\Large$\bm{(0,0)}$} at -29 10 
\put {\Large$\bm{(0,2)}$} at -29 50 


\setplotsymbol ({$\cdot$})
\circulararc 360 degrees from 0 8 center at 0 0 

\normalcolor\color{black}

\arrow <10pt> [.2,.67] from 0 -15 to 240 -15
\multiput {\scalebox{0.75}{$\bullet$}} at 0 -15 / 
\put {\LARGE $w$} at 120 -25

\endpicture
\caption{Two osculating paths from vertices with coordinates $(0,0)$ and
$(0,2)$ in the square lattice, giving NE and SE steps, and ending in vertices
with coordinates $(n,h_1)$ and $(n,h_2)$, and heights $h_1<h_2$.  Returns 
of the bottom path to the hard wall (the $x$-axis) are weighted by the 
parameter $a$.  The width $w$ of the paths is indicated and 
is the number of steps in bottom path.  This is also the half-length 
of the pair of paths.}
\label{F3}
\end{figure}

The left-most and right-most two steps in a staircase polygon have fixed
orientation and may be removed without affecting the combinatorial
properties of the remaining paths.  This gives two directed paths (a bottom 
path and a top path).  More generally, these paths may be allowed to end 
in vertices which are farther apart, so that they cannot be reconnected to 
form a staircase polygon.  This is illustrated in figure \ref{F3}.  Since 
the paths avoid one another, but may approach closely to ``bounce" off 
each other, they are also called \textit{osculating paths}; see for 
example reference \cite{BM06}.

Consider vertices $(n,m)$ of even parity (that is, $n+m$ is even) in $\mathL$.
The bottom directed path in figure \ref{F3} starts in $(0,0)$, has length $n$,
and ends in $(n,h_1)$, while the top path starts in $(0,2)$ and ends in
$(n,h_2)$.  The width of the pair of paths is $w = n$.  Two steps can be
prepended to the left in figure \ref{F3}, and if $h_2=h_1+2$, then 
another two steps can be appended to the right to join the two paths
into a staircase polygon.  More generally, if only the two steps are 
prepended to the left in figure \ref{F3}, then the two paths is a 
directed path model of a $2$-star copolymer (figure \ref{F1}(left)), with 
its bottom arm adsorbing in a \textit{hard wall}.

If $h_1=0$ and $h_2=2$ in figure \ref{F3}, and steps are prepended to the
left and appended on the right, the paths become a 
staircase polygon grafted at both endpoints (so that the second and 
penultimate vertices in the bottom path are grafted to the hard wall).
This staircase polygon have total length $2n+4$ and width $w=n$.  Returns 
of the bottom path to the hard wall, except at the grafted vertices, are 
weighted by $a$.  If $a=1$, then the bottom path is said to be free, and 
the paths or staircase polygon is in a \textit{desorbed or free phase} 
\cite{JvRP13}.   For $a>2$ the bottom path is adsorbed on the wall, and 
the model is in an \textit{adsorbed phase}.  These phases are separated 
by the adsorption critical point $a_c=2$ (the \textit{special point});  for 
more results about the phase diagram of adsorbing staircase polygons, 
see reference \cite{JvR15}.

\begin{figure}[t]
\begin{center}
\includegraphics[width=0.66\textwidth]{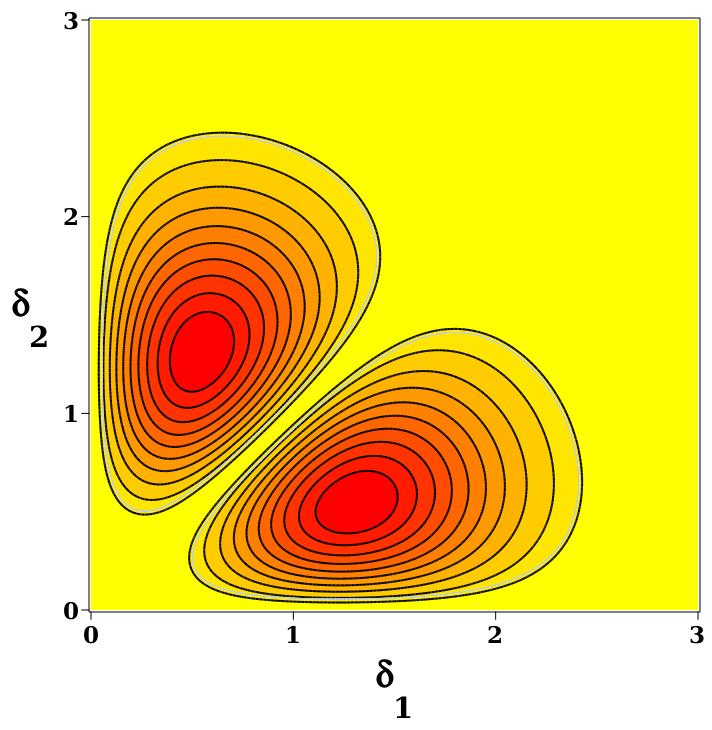}
\end{center}
\caption{Contourplot of $|\mathP_r^{(2)} (1,\delta_1,\delta_2)|$
on the $\delta_1\delta_2$-plane.  Low values are in light shades, and 
high values are in darker shades.  This shows that paths with 
endpoints at $(\delta_1,\delta_2) \approx(0.5412,1.3066)$ dominate 
the partition function in the limit as $n\to\infty$, so that the most 
likely heights of the endpoints are approximately 
$h_1\approx 0.5412\,\sqrt{n}$ and $h_2\approx 1.3066\,\sqrt{n}$.}
\label{F4}
\end{figure}

\subsection{Probability densities of grafted staircase polygons}
\label{S1.2}

The paths in figures \ref{F2} and \ref{F3} have length $n$ and are ballistic 
in the horizontal direction so that the natural horizontal length scale is 
$O(n)$.  In the vertical direction the paths are random walks, with vertical 
length scale $O(\sqrt{n})$.  By rescaling the lattice 
an asymptotic probability distribution for paths passing through 
specific lattice sites in $\IntZ^2$ can be determined.  Though not rigorous, 
this distribution converges in the limit as $n\to\infty$ to the (exact) 
probability density of the paths passing through points in $\RealN^2$.  
This is the \textit{scaling limit} of the model.  The calculation relies on 
(non-rigorous) asymptotic estimates of the finite size probability
distribution.

In section \ref{S2} the model of osculating paths in figure \ref{F3} is 
considered in the free phase (with $a=1$).  The number of pairs of paths 
of width $n$ and heights $h_1=j-\ell$ and $h_2=j+\ell+2$  is given 
by (see for example reference \cite{JvR15} equation (5.172))
\begin{equation}
\hspace{-1.0cm}
\hbox{\scalebox{1.0}{$\dps
U_n(j,\ell) = \frac{
(j{+}2)\,
(\ell{+}1)\,
(j{-}\ell{+}1)\,
(j{+}\ell{+}3)}{
(n{+}1)\,
(n{+}2)\,
(n{+}3)^2} \;
\Bi{n+3}{(n{+}j{-}\ell)/2+2)}\Bi{n+3}{(n{+}j{+}\ell)/2+3)}$}},
\label{eqn1}
\end{equation}
where $j$ and $\ell$ have the same parity as $n$ and $j\geq\ell$.  By putting
$j=\ell=0$ and  ``capping'' the left starting points the number of staircase 
polygon of length $2n+4$, grafted at $(0,0)$ and $(n,0)$, is given by
by $U_n(0,0)$.

Define $V_n(h_1,h_2) \equiv U_n(j,\ell)$ (with $h_1=j-\ell$ and $h_2=j+\ell+2$). 
A probability distribution can be determined from $U_n(j,\ell)$ for the probability
that the two paths end in heights $h_1=\lfl \delta_1 \sqrt{n}\rfl$ and 
$h_2=\lfl \delta_2 \sqrt{n} \rfl$.  This is obtained from equation \Ref{eqn1}
by subsituting $h_1=j-\ell=\lfl \delta_1 \sqrt{n}\rfl$ and 
$h_2=j+\ell+2=\lfl \delta_2 \sqrt{n}\rfl$, and then normalising.  To 
leading order,
\begin{equation}
U_{\lfl \sigma n\rfl} 
= V_{\lfl \sigma n\rfl} 
(\lfl \delta_1\sqrt{n} \rfl,\lfl \delta_2\sqrt{n}\rfl)
 \equiv V_{\lfl \sigma n\rfl} (\delta_1,\delta_2)
\simeq \frac{\delta_1\delta_2(\delta_2^2-\delta_1^2)}{\pi n^3\sigma^5}
\, 16^{n\sigma + 1}\, e^{-(\delta_1^2+\delta_2^2)/\sigma} ,
\label{eqn2A}
\end{equation}
where the arguments of $U_{\lfl \sigma n\rfl} (j,\ell)$ are surpressed for clarity.
The normalising factor $N_u$ for $U_n$ is obtained by integrating it for 
$0< \delta_1 < \delta_2 < \infty$.  This gives
\begin{equation}
N_u = \int_0^\infty \int_{\delta_1}^\infty U_n\,d\delta_2\,d\delta_1
= \frac{16^{\sigma n + 1/4}}{\pi n^3 \sigma^2}.
\end{equation}
Dividing $U_{\lfl\sigma n\rfl}/N_u$ gives the asymptotic probability density
which is independent of $n$ and therefore is the exact probability
probability density in the scaling limit:
\begin{equation}
\mathP_r^{(2)} (\sigma,\delta_1,\delta_2) 
= \frac{8\,\delta_1\delta_2(\delta_2^2-\delta_1^2)}{\sigma^3}\;
e^{-(\delta_1^2+\delta_2^2)/\sigma} ,
\quad\hbox{for $0\leq \delta_1 \leq \delta_2 <\infty$}.
\label{eqn2}
\end{equation} 
Details are given in section \ref{S2}.  

Putting $\sigma=1$ and plotting the absolute value 
$|\mathP_r^{(2)} (1,\delta_1,\delta_2)|$ in the $\delta_1\delta_2$-plane 
gives figure \ref{F4}.  Notice that this is the asymptotic probability 
density that the endpoints of paths of length $n$ end in 
the vertices $(n,\lfl\delta_1\sqrt{n}\rfl)$ (bottom path), and 
$(n,\lfl\delta_2\sqrt{n}\rfl)$ (top path) \textit{in the scaling limit}.  
By determining the maxima in the plot, the most likely heights of 
the endpoints are obtained with 
$\delta_1=\sqrt{\sqrt{2}-1}/\sqrt{8} \approx 0.5412$
and $\delta_2=\sqrt{\sqrt{2}+2}/\sqrt{2} \approx 1.3066$.

\begin{figure}[t]
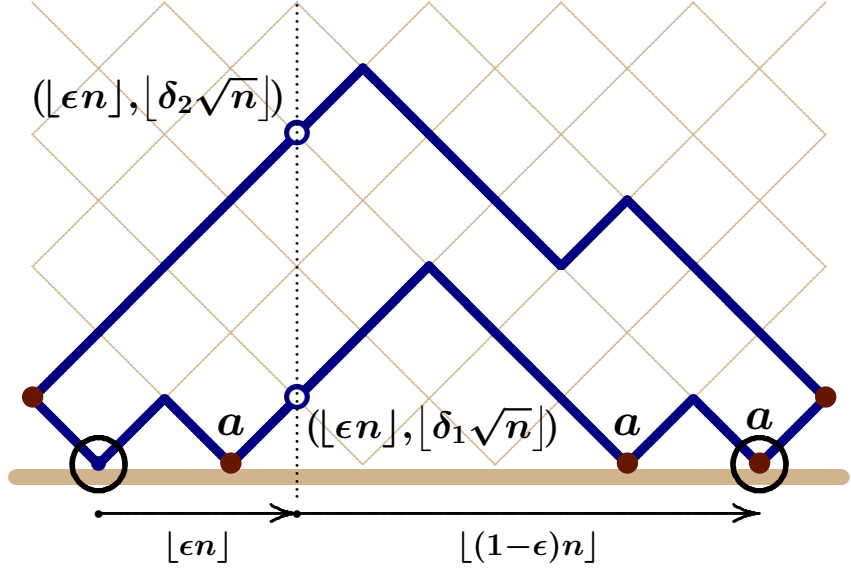

\centering
\beginpicture
\setcoordinatesystem units <1.25pt,1.25pt>
\setplotarea x from -50 to 220, y from -15 to 140

\color{Tan}
\plot 0 40 100 140 / \plot 0 80 60 140 / \plot 0 120 20 140 /
\plot 0 0 140 140 / \plot 40 0 180 140 / \plot 80 0 220 140 / 
\plot 120 0 220 100 / \plot 160 0 220 60 / \plot 200 0 220 20 / 

\plot 40 0 0 40 / \plot 80 0 0 80 / \plot 120 0 0 120 / \plot 160 0 20 140 /
\plot 200 0 60 140 / \plot 220 20 100 140 / \plot 220 60 140 140 /
\plot 220 100 180 140 /

\plot 0 0 -20 20 0 40 -20 60 0 80 -20 100 0 120 -20 140 /

\color{NavyBlue}
\setplotsymbol ({\footnotesize$\bullet$})

\plot -20 20 0 40 20 60 40 80 60 100 80 120 100 100 120 80 140 60 160 80 180 60 200 40 220 20 /
\plot -20 20 0 0 20 20 40 0 60 20 80 40 100 60 120 40 140 20 160 0 180 20 200 0 220 20 /

\setplotsymbol ({\scalebox{1.5}{$\bullet$}})
\color{Tan}
\plot -25 -4 225 -4 /

\color{Sepia}
\multiput {\huge$\bullet$} at -20 20 40 0 160 0 200 0 220 20  / 
\color{NavyBlue}
\multiput {\Large$\bullet$} at 0 0 / 
\multiput {\scalebox{2.5}{$\bullet$}} at 60 100 / 
\multiput {\scalebox{2.5}{$\bullet$}} at 60 20 / 
\color{White}
\multiput {\scalebox{1.5}{$\bullet$}} at 60 100 / 
\multiput {\scalebox{1.5}{$\bullet$}} at 60 20 / 

\color{Black}
\multiput {\LARGE$\bm{a}$} at 40 12 160 12 200 14 /

\put{\Large$\bm{(\lfl\eps\thin n\rfl,\lfl \delta_2 \sqrt{n}\rfl)}$} at 18 111
\put{\Large$\bm{(\lfl\eps\thin n\rfl,\lfl \delta_1 \sqrt{n}\rfl)}$} at 101 11

\setplotsymbol ({\scalebox{0.5}{$\bullet$}})
\circulararc 360 degrees from 0 8 center at 0 0 
\circulararc 360 degrees from 200 8 center at 200 0 

\normalcolor\color{black}

\setplotsymbol ({$\cdot$})
\arrow <10pt> [.2,.67] from 0 -15 to 60 -15
\put {\large$\bm{\lfl\eps n\rfl}$} at 30 -25
\arrow <10pt> [.2,.67] from 60 -15 to 200 -15
\multiput {\scalebox{0.75}{$\bullet$}} at 0 -15 60 -15 / 
\put {\large$\bm{\lfl(1{-}\eps) n\rfl}$} at 130 -25

\setdots <3pt>
\plot 60 -10 60 140 /

\endpicture
\caption{A staircase polygon grafted at the origin and at the end of its bottom
directed path.  By cutting the path along the vertical dotted line, two pairs of
directed paths, one pair from the left, and the other from the right, are obtained.
The probability that the bottom path passes through the vertex with coordinates
$(\lfl\eps n\rfl,\lfl \delta_1 n\rfl)$, and the top path passes through the vertex
with coordinates $(\lfl\eps n\rfl,\lfl \delta_2 n\rfl)$ is obtained by approximating
the number of pairs of paths passing through these vertices.}
\label{F5}
\end{figure}

The asymptotic probability that the staircase polygon passes through a vertex 
$(\lfl \eps n\rfl,\lfl \delta \sqrt{n}\rfl)$ can be determined from equation 
\Ref{eqn1}, and its asymptotic form $V_{\lfl \sigma n\rfl} (\delta_1,\delta_2)$ 
in equation \Ref{eqn2A}.  In figure \ref{F5} a grafted staircase polygon
passing through the points $(\lfl\eps\thin n\rfl,\lfl \delta_1 \sqrt{n}\rfl)$
(in its bottom branch) and $(\lfl\eps\thin n\rfl,\lfl \delta_2 \sqrt{n}\rfl)$ 
(in its top branch) is illustrated.  The vertical 
dotted line splits the polygon into two parts, one part a pair of osculating paths 
of width $\lfl \eps n \rfl$ from the left, and the other a pair of osculating 
paths of width $\lfl (1{-}\eps) n \rfl$ from the right.

\begin{figure}[t]
\begin{center}
\includegraphics[width=0.66\textwidth]{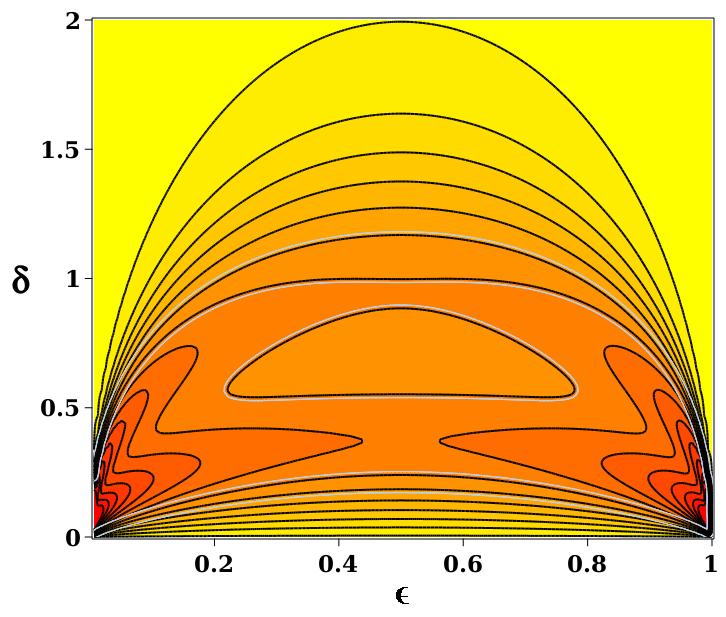}
\end{center}
\caption{Contourplot of the density $\mathP^{f}_r (\eps,\delta)$
of grafted staircase polygons in the scaling limit.  Lighter regions have 
low densities.  The plot shows low densities adjacent to the hard wall, 
and also a region of relative low density between the top and 
bottom paths.}
\label{F6}
\end{figure}

The asymptotic number of pairs of osculating paths are given in equation
\Ref{eqn2A}. This gives the asymptotic number of staircase polygons in
figure \ref{F5} as approximately
\begin{equation}
V_{\lfl \eps n\rfl}\LB \lfl \delta_1 \sqrt{n} \rfl,\lfl \delta_2 \sqrt{n} \rfl \RB
\times
V_{\lfl (1{-}\eps) n\rfl}\LB \lfl \delta_1 \sqrt{n} \rfl,\lfl \delta_2 \sqrt{n} \rfl \RB .
\end{equation}
By normalising this expression, the probability density that the bottom
path passes through the vertex with coordinates 
$(\lfl \eps n \rfl, \lfl \delta_1 \sqrt{n} \rfl)$, and the top path 
passes through $(\lfl \eps n \rfl, \lfl \delta_2 \sqrt{n} \rfl)$, is obtained
in the scaling limit.  This is given by
\begin{equation}
\mathP^{f}_r (\eps,\delta_1,\delta_2) 
= \frac{32}{3}\,
\frac{(\delta_2^2-\delta_1^2)^2\,\delta_1^2\delta_2^2}{
\pi\,\eps^5\,(1{-}\eps)^5}\, e^{-(\delta_1^2+\delta_2^2)/\eps(1{-}\eps)}.
\label{eqn6A}
\end{equation}
Integrating $\mathP^{f}_r(\eps,\delta_1,\delta_2)$ for
$0 < \delta_2 < \infty$ and putting $\delta_1=\delta$ gives the probability 
density 
\begin{equation}
\mathP^{f,(both)}_r (\eps,\delta)
= \int_0^\infty \mathP^{(2)}_r (\eps,\delta,\delta_2)\, d\delta_2
\end{equation} 
that either the bottom path, or the top path, passes through points 
$(\eps,\delta)$ in the infinite rectangle $S=[0,1]\times[0,\infty) 
\subseteq \RealN^2$. Simplifying the result gives
\begin{equation}
\mathP^{f,(both)}_r (\eps,\delta)
= 
\frac{\delta^2(15\,\eps^2(1{-}\eps)^2
-12\,\delta^2\eps(1{-}\eps)
+4\,\delta^4)}{
3\,\sqrt{\pi}\,\sqrt{\eps^7(1{-}\eps)^7}}
\, e^{-\delta^2/\eps(1{-}\eps)} .
\label{eqn8A}
\end{equation}
A contourplot of this density is plotted in figure \ref{F6}.  Notice that there 
are low density regions next to the hard wall and between the trajectories 
of the bottom and top paths.  This shows that the entropic repulsion between
the paths, and between the bottom path and the wall, create low density
regions in this model of polymer coatings that particles can occupy in
a metastable state before they are released when overcoming 
the entropic barriers.

The domain of the probability density $\mathP^{f,(both)}_r (\eps,\delta)$
can be extended from $S$ to all of $\RealN^2$ by defining
$\mathP^{f,(both)}_r (\eps,\delta) = 0$ if $(\eps,\delta)\not\in S$.
Define $R = (x_1,x_2)\times(y_1,y_2)$ to be an open rectangle in 
$\RealN^2$, and suppose that $\o{R}$ is the closure of $R$.  Define 
the set function $\nu$ by
\begin{equation}
\nu R = \nu \o{R} = \int_{-\infty}^{\infty} \int_{-\infty}^{\infty} 
  \mathP^{f,(both)}_r (\eps,\delta)\, d\delta\,d\eps .
\label{eqn9A}
\end{equation}
Then $\nu$ is a set-function on the semi-algebra of rectangles in
$\RealN^2$, and it is additive on unions of disjoint rectangles.
It follows that $\nu (R\cap S) = \nu R$, and on subsets of the 
$\eps\delta$-plane $\nu$ extends to an outer measure.  On the 
$\sigma$-algebra of plane measure sets $\nu$ is a measure, and
if $E$ is a plane measurable set with respect to plane measure $\lambda$, 
then
\begin{equation}
\nu E = \int_E \mathP^{f,(both)}_r (\eps,\delta)\, d\lambda .
\label{eqn10A}
\end{equation}
The Radon-Nikodym derivative of $\nu$ with respect to $\lambda$ is 
$\frac{d\nu}{d\lambda} = \mathP^{f,(both)}_r (\eps,\delta)$ at the 
point $(\eps,\delta)\in\RealN^2$.  Since $\nu \RealN^2 = 1$,  it is a 
probability measure on $\lambda$-measure sets, and $\nu E$ is the 
probability that paths from the origin $(0,0)$ to $(1,0)$ pass through $E$.  

\subsection{Organization of the paper}

In section \ref{S2} some details of the derivation of equation \Ref{eqn2A} are
given.  The most probable trajectories of the bottom and top paths are 
determined, as well as the probability densities of the bottom path, or the
top path, passing through a point, and their most likely trajectories.

In section \ref{S3} the cases of adsorbed staircase polygons are considered.
The situation is more complicated and some details are given in the Appendix.
In this case the bottom path is adsorbed on the wall,  changing the density
of the model.  The density at the critical point $a_c=2$ (the 
\textit{special point}) is also distinct from either the densities in the free 
or in the adsorbed phase, and this is presented in section \ref{S4}.

A final discussion of the results, as well as the most likely and mean 
trajectories and entropic forces, are given in section \ref{S5}.

\section{The densities of paths in the free phase}
\label{S2}

\subsection{The density of both paths in the free phase}

In this section more details of the derivation of equations \Ref{eqn2A} 
and \Ref{eqn2} are given.  Substituting and approximating 
$j-\ell=h_1=\lfl \delta_1\sqrt{n}\rfl\simeq \delta_1\sqrt{n}$, 
$j+\ell+2=h_2=\lfl\delta_2\sqrt{n}\rfl\simeq \delta_2\sqrt{n}$ 
and replacing and then approximating 
$n \to \lfl \sigma n \rfl \simeq \sigma n$ in equation \Ref{eqn2A} gives
(where the arguments of $V_{\lfl \sigma n \rfl}$ are left away for 
simplicity)
\begin{equation*}
\hspace{-0.5cm}
V_{\lfl \sigma n \rfl} 
=\hbox{\scalebox{0.85}{$\displaystyle
\frac{((\delta_2{-}\delta_1)\sqrt{n})\,
((\delta_1{+}\delta_2)\sqrt{n}{+}1)\,
(2\,\delta_1\sqrt{n}{+}1)\,
(2\,\delta_2\sqrt{n}{+}1)
}{(2\,n\sigma{+}1)\,
(2\,n\sigma{+}2)\,
(2\,n\sigma{+}3)^2}\;
\Bi{2\,n\sigma+3}{n\sigma+\delta_1\sqrt{n}+2} 
\Bi{2\,n\sigma+3}{n\sigma+\delta_2\sqrt{n}+2} $}}.
\end{equation*}
Convert the binomial coefficients into factorials.  Using the Stirling 
approximation in equation \Ref{eqn56A} to approximate the factorials 
and simplifying the result shows that, to leading order,
\begin{equation}
V_{\lfl \sigma n\rfl} \simeq
\frac{3}{2\pi}
\frac{\sqrt{n}\,(\delta_2-\delta_1)\,
a_0a_1a_3a_4^{a_{16}}a_5a_{11}^{a_{18}}
}{\sqrt{a_7a_8a_9a_{10}}\, a_6^{a_{17}}a_{12}^{a_{19}}a_{13}^{a_2}
\, a_{14} a_{15} } ,
\label{eqn8}
\end{equation}
where
\begin{align*}
a_0&=\delta_1\sqrt{n} +\delta_2\sqrt{n} +1, &
a_1&=2\,\delta_1\sqrt{n} +1, &
a_2&=n\sigma - \delta_2\sqrt{n} +1 \\
a_3&=2\,\delta_2\sqrt{n} +1, &
a_4&=2\,n\sigma +3, &
a_5&=12\,n\sigma +19, \\
a_6&=n\sigma +\delta_1\sqrt{n} +2, &
a_7&=6\,n\sigma +6\,\delta_1\sqrt{n} +13, &
a_8&=6\,n\sigma -6\,\delta_1\sqrt{n} +7, \\
a_9&=6\,n\sigma +6\,\delta_2\sqrt{n} +13, &
a_{10}&=6\,n\sigma -6\,\delta_2\sqrt{n} +7, &
a_{11}&=n\sigma -\delta_1\sqrt{n} +1 , \\
a_{12}&=n\sigma +\delta_2\sqrt{n} +2, &
a_{13}&=n\sigma -\delta_2\sqrt{n}+1 , &
a_{14}&=2\,n\sigma +1, \\
a_{15}&=n\sigma +1, &
a_{16}&=4\,n\sigma +4, &
a_{17}&=n\sigma +\delta_1\sqrt{n} +2, \\
a_{18}&=\delta_1\sqrt{n} -n\sigma -1, &
a_{19}&=n\sigma +\delta_2\sqrt{n} +2 .
\end{align*}
Taking the logarithm of equation \Ref{eqn8} and expanding, and then 
expanding the resulting terms  each asymptotically while discarding 
terms decaying with $n$ give, after exponentiating and simplifying 
the result, equation \Ref{eqn2A}, and, after normalisation,
the probability density in equation \Ref{eqn2}.

\subsubsection{The most likely and mean trajectories in the free phase:}
The density plot in figure \ref{F6} shows that there are \textit{most likely
trajectories} for the bottom and top paths.  These may be assumed to run 
along the peaks in the plot so that they are ``modal paths" from $(0,0)$ to 
$(1,0)$. Solving for local minima and maxima for fixed values of $\eps\in (0,1)$
in equation \Ref{eqn6A} shows that the extreme values of $\delta$ as a 
function of $\eps$ is given by solutions of the polynomial
\begin{equation}
8\,\delta^6 - 48\,\delta^4\,\eps(1{-}\eps) 
  + 78\,\delta^2\eps^2(1{-}\eps)^2 
    - 30\,\eps^3(1{-}\eps)^3 = 0.
\end{equation}
Solving for $\delta$ gives $\delta_M$, the most likely trajectories of the
paths, and also its (locally) least likely trajectory (the ``valley'' between the top
and bottom paths):
\begin{numcases}{\delta_M^2 =}
(\sqrt{3}/2){\Large\hbox{$[$}}
        \sin((\arctan\sqrt{26})/3 + \pi/6) \nonumber \\
 \hspace{1cm} + \sqrt{3} \sin((-\arctan\sqrt{26})/3 + \pi/3) 
                         + 4\sqrt{3} {\Large\hbox{$]$}} \,{\eps(1{-}\eps)}, 
                                                                                & \hbox{top trajectory} ; 
                                                                                  \nonumber \\
(\sqrt{3}/2){\Large\hbox{$[$}}
        \sin((\arctan\sqrt{26})/3 + \pi/6) \label{eqn12Q} \\
 \hspace{1cm} - \sqrt{3} \sin((-\arctan\sqrt{26})/3 + \pi/3) 
                 + 4\sqrt{3} {\Large\hbox{$]$}} \,{\eps(1{-}\eps)}, & \hbox{valley} ;
                 \nonumber \\
(2-\sqrt{3}\sin((\arctan\sqrt{26})/3 
                 + \pi/6))\,{\eps(1{-}\eps)}, & \hbox{bottom trajectory} . \nonumber 
\end{numcases}
Evaluating the constants gives
\begin{equation}
\delta_M =
\begin{cases}
1.8848\ldots\sqrt{\eps(1{-}\eps)}, & \hbox{top trajectory} ; \\
1.3741\ldots\sqrt{\eps(1{-}\eps)}, & \hbox{valley} ; \\
 0.7476\ldots\sqrt{\eps(1{-}\eps)}, & \hbox{bottom trajectory} . 
\end{cases}
\label{eqn13A}
\end{equation}
The mean path can also be determined by integrating 
$\delta\,\mathP_r^{f,(both)}(\eps,\delta)$.  This gives the mean path
$\o{\delta} = (5/2\,\sqrt{\pi})\,\sqrt{\eps(1{-}\eps)}$ for the full model.

\begin{figure}[t]
\begin{center}
\includegraphics[width=0.48\textwidth]{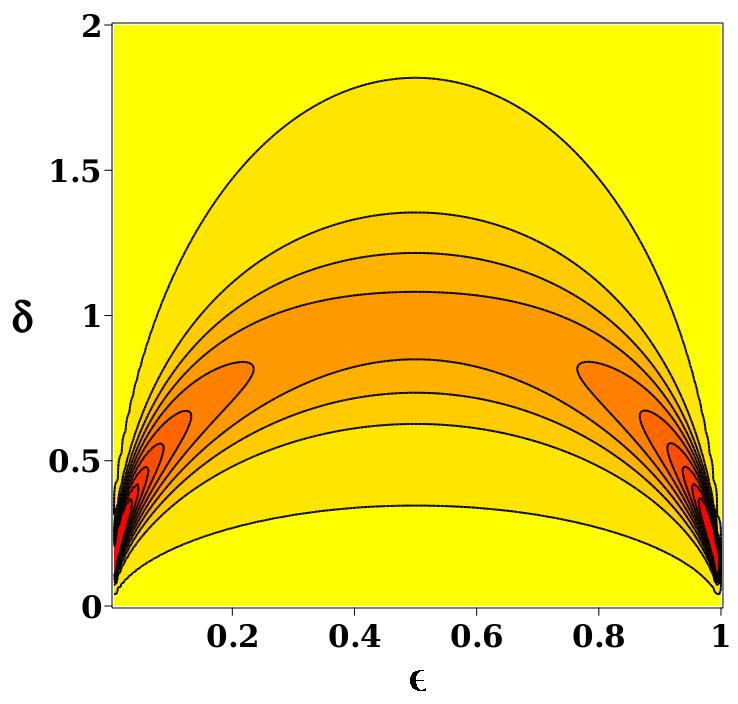}
\includegraphics[width=0.48\textwidth]{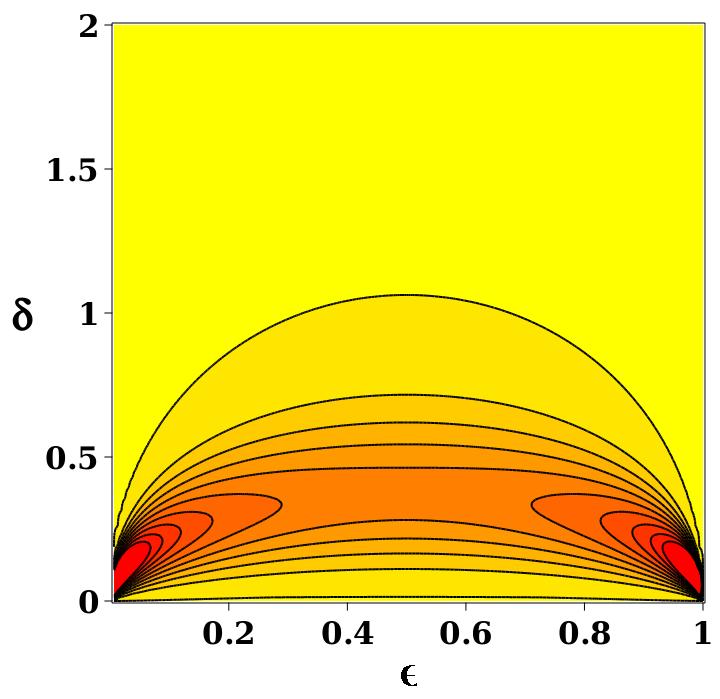}
\end{center}
\caption{(Left) Contourplot of the density $\mathP^{f,(top)}_r (\eps,\delta)$
of the top path in grafted staircase polygons in the scaling limit.  
(Right) A contourplot of the density $\mathP^{f,(bottom)}_r (\eps,\delta)$ 
of the bottom path in grafted staircase polygons in the scaling limit.
Lighter regions have low densities.}
\label{F7}
\end{figure}

\subsection{The probability density of the top path in the free phase}

The probability that the top path passes through a given point $(\eps,\delta)$
can be obtained by integrating $\mathP_r^{(2)}(\eps,\delta_1,\delta_2)$
(equation \Ref{eqn2}) for $\delta_1\in (0,\delta_2)$ and then normalising
the result.  Putting $\delta_2=\delta$ gives
\begin{align}
\mathP_r^{f,(top)} (\eps,\delta)
&= \frac{2}{3} \,
\frac{\delta^2(4\,\delta^4-12\,\delta^2\eps(1{-}\eps) + 15\,\eps^2(1{-}\eps)^2)
}{\sqrt{\pi}\,\sqrt{\eps^7(1{-}\eps)^7}}\,e^{-\delta^2/\eps(1{-}\eps)}
\,\erf(\delta/\sqrt{\eps(1{-}\eps)})  \nonumber \\
&\hspace{2cm}
 - \frac{4}{3}\, \frac{\delta^3(2\,\delta^2-15\,\eps(1{-}\eps))}{
            \pi\,\eps^3(1{-}\eps)^3}\,e^{-2\,\delta^2/\eps(1{-}\eps)}  .
\label{15Q}
\end{align}
A plot of the density of the top path is shown in figure \ref{F7} (left panel).
A probality measure $\nu_t$ with respect to plane measure can be defined 
as in equation \Ref{eqn10A} for calculating the probabilities that the top
path passes through a subset in the plane, irrespective of the trajectory of
the bottom path.

\subsubsection{The most likely and mean paths of the top path in the free phase:}
Taking the derivative of $\log \mathP_r^{(top)}(\eps,\delta)$ 
(equation \Ref{15Q}), and determining its numerator, show that 
the most likely trajectory of the top path is determined by 
the solution of $\delta$ in 
\begin{align}
&4\sqrt{\pi}\,\delta \sqrt{\eps(1{-}\eps)}
\LB 2\,\delta^4 -29\,\delta^2\eps(1{-}\eps)+15\,\eps^2(1{-}\eps)^2 \RB 
e^{-\delta^2/\eps(1{-}\eps)} \nonumber \\
&+ 2\,\pi \LB 4\,\delta^6 - 24\, \delta^4\eps(1{-}\eps) 
      + 39\,\delta^2\eps^2(1{-}\eps)^2 - 15\,\eps^3(1{-}\eps)^3 \RB 
          \erf(\delta/\sqrt{\eps(1{-}\eps)}) = 0 .
\end{align}
Simplify this by putting $\delta^2=K\,\eps(1{-}\eps)$.  This gives
\begin{equation}
4\sqrt{\pi}\,(2\,K^2-29\,K+15)\,e^{-K} +
2\,\pi\,(4\,K^3 - 24\, K^2 + 39\, K - 15)\, \erf(\sqrt{K}) = 0  
\label{eqn16A}
\end{equation}
which can be solved numerically.  Choosing the correct solution
gives $K=1.9082\ldots$.  This gives the most likely 
trajectory $\delta_M = 1.3814...\sqrt{\eps(1{-}\eps)}$. This 
is not the same trajectory of the top path in equation \Ref{eqn13A} 
(since integrating the bottom path changes the weight 
of the top path in the probability distribution).

The mean top path can be calculated from $\mathP_r^{f,(top)}(\eps,\delta)$ 
and is given by $\o{\delta} = (5/\sqrt{2\pi}) \sqrt{\eps(1{-}\eps)}$, which
is larger by a factor of $\sqrt{2}$ than the mean trajectory of both
paths.

\subsection{The probability density of the bottom path in the free phase}

The density of the bottom path is obtained by integrating equation 
\Ref{eqn8A} for $\delta_2\in(\delta_1,\infty)$ and then normalising.  
This gives 
\begin{align}
\mathP_r^{f,(bottom)} (\eps,\delta)
&= \frac{2}{3} 
\frac{\delta^2(4\,\delta^4-12\,\delta^2\eps(1{-}\eps) + 15\,\eps^2(1{-}\eps)^2)
}{\sqrt{\pi}\,\sqrt{\eps^7(1{-}\eps)^7}}\,e^{-\delta^2/\eps(1{-}\eps)}
\,\erfc(\delta/\sqrt{\eps(1{-}\eps)})  \nonumber \\
&\hspace{2cm}
 + \frac{4}{3} \frac{\delta^3(2\,\delta^2-15\,\eps(1{-}\eps))
}{\pi\,\eps^3(1{-}\eps)^3}\,e^{-2\,\delta^2/\eps(1{-}\eps)}  ,
\end{align}
where $\erfc(x) = 1-\erf(x)$ is the complementary error function.  
This density is plotted in the right panel of figure \ref{F7}.  Notice that
\begin{equation}
2\,\mathP^{f,(both)}_r(\eps,\delta)
= \mathP^{f,(top)}_r(\eps,\delta) +\mathP^{f,(bottom)}_r(\eps,\delta) ,
\end{equation}
as expected.

A probality measure $\nu_b$ with respect to plane measure can be defined 
as in equation \Ref{eqn10A} for calculating the probabilities that the bottom
path passes through a subset in the plane, irrespective of the trajectory of
the top path.

\subsubsection{The most likely and mean paths of the bottom path in the free phase:} To determine the most likely trajectory, proceed as above by 
assuming that it is a curve $\delta = K\sqrt{\eps(1{-}\eps)}$.  This 
again gives equation \Ref{eqn16A} which can be solved numerically 
and choosing the correct zero.  This gives $K=0.7073\ldots$ and 
notice that $K^2=0.500232\ldots$, very close to $1/2$.  The most
likely trajectory is $\delta_M = 0.7073\ldots\sqrt{\eps(1{-}\eps)}$.

The mean trajectory of the bottom path can be calculated from 
$\mathP_r^{f,(bottom)} (\eps,\delta)$. This gives 
$\o{\delta} = (5(\sqrt{2}{-}1)/\sqrt{2\pi}) \sqrt{\eps(1{-}\eps)}$.

\section{Probability densities in the adsorbed phase}
\label{S3}

The number $r_n(j,\ell)$ of pairs of osculating paths (see figure \ref{F3}), 
starting in $(0,0)$ and $(0,2)$, each of length $n$ (so that the width of 
the pair of paths is $n$), and ending in vertices with heights $h_1=(j-\ell)$ 
and $h_2=(j+\ell+2)$, is given by equation (5.173) in reference \cite{JvR15}:
\begin{equation}
r_n(j,\ell) = \sum_{m_1=0}^n \sum_{m_2=0}^{n+1}
U_{n}(j+m_1+m_2,\ell+m_1+m_2)\, (a-1)^{m_1+m_2}
\label{A1}
\end{equation}
where $a$ is the generating variable of returns of the 
bottom path to the adsorbing hard wall.  $U_n(j,\ell)$ is given by 
equation \Ref{eqn1}, and $j$ and $\ell$ must have the same parity 
as $n$.   In this paper the expressions are simplified by putting 
$A=a-1$, so that the special or critical point is $A=1$ (since $a_c=2$
is the location of the critical points).

If $\ell=0$ then these are adsorbing directed paths from $(0,0)$ and $(0,2)$ 
to the vertices $(n,j)$ and $(n,j+2)$.  By ``capping'' the left starting points 
with two steps, and the right end points with two steps, an adsorbing 
staircase polygon of length $2n+4$,  grafted at only at the origin, is obtained.  
The partition function of such adsorbing polygons is given by $r_n(j,0)$.  
Similarly, if $j=\ell=0$, then the partition function $r_n(0,0)$ is of a model
of adsorbing staircase polygons grafted at both $(0,0)$ and $(n,0)$.  
These polygons have length $2n+4$ and are grafted at both ends as 
shown in figure \ref{F5}. For example, if $n=4$ then
\begin{equation}
r_n(0,0) = 14\,a^3 + 28\,a^2 + 28\,a + 14 .
\end{equation}
Define 
\begin{equation}
R_n(h_1,h_2) = r_n((h_1{+}h_2)/2-1,(h_2{-}h_1)/2-1),
\label{eqn19A}
\end{equation}
the partition function of pairs of paths from $(0,0)$ and $(0,2)$ respectively,
and to $(n,h_1)$ and $(n,h_2)$ respectively.

The asymptotic behaviour of $R_n(h_1,h_2)$ is determined in the Appendix,
and it is given by equation \Ref{eqnA10} when $a>2$ (recall that $a=A{+}1$).
It may be put into the form
\begin{equation}
R_{2\,\lfl \sigma n\rfl}(2\,\lfl \delta_1\sqrt{n}\rfl,2\,\lfl \delta_2\sqrt{n}\rfl)
\simeq F(\sigma)\, 
\LB \delta_2 \, A^{-\delta_1\sqrt{n}}\,e^{-\delta_2^2/\sigma}
 - \delta_1\, A^{-\delta_2\sqrt{n}}\,e^{-\delta_1^2/\sigma} \RB,
\label{eqn20A}
\end{equation}
where the approximations $\lfl \sigma n \rfl \simeq \sigma n$,
$\lfl \delta_1\sqrt{n}\rfl \simeq \delta_1\sqrt{n}$ and 
$\lfl \delta_2\sqrt{n}\rfl \simeq \delta_2\sqrt{n}$ are made to obtain
\begin{equation}
F(\sigma) = \frac{2 \log^2 A}{\sqrt{\pi \sigma^3 n^3}}\,
16^{\sigma n} e^{\sigma n (\log A)^2/4} .
\end{equation}
The probability that staircase polygons grafted at $(0,0)$ and $(2n,0)$ and of
width $2n$ pass through points with coordinates 
$(2\lfl \eps n \rfl,2\lfl \delta_1\sqrt{n}\rfl)$ 
(the bottom path) and $(2\lfl \eps n \rfl,2\lfl \delta_2\sqrt{n}\rfl)$ (the top path)
is asymptotically approximated by 
\begin{equation}
N_o\,R_{2\, \lfl \eps n\rfl}(2\,\lfl \delta_1\sqrt{n}\rfl,2\,\lfl \delta_2\sqrt{n}\rfl)
\times 
R_{2\,\lfl (1{-}\eps) n\rfl}(2\,\lfl \delta_1\sqrt{n}\rfl,2\,\lfl \delta_2\sqrt{n}\rfl),
\label{eqnA22}
\end{equation}
where $N_o$ is a normalising factor, independent of $(\eps,\delta_1,\delta_2)$.
Using the approximation in equation \Ref{eqn20A} and then expanding the
result gives 
\begin{eqnarray}
&
\hspace{-2.5cm}
P_n^{A}(\eps,\delta_1,\delta_2)
\simeq N_o\,F(\eps)F(1{-}\eps)\,
\LB
   A^{-2\,\delta_1\sqrt{n}}\,\delta_2^2\,e^{-\delta_2^2/\eps(1{-}\eps)}
   - A^{-(\delta_1+\delta_2)\sqrt{n}}\, \delta_1\delta_2\, e^{-\delta_2^2/\eps - \delta_1^2/(1{-}\eps)}
\right. \cr
& \left.  
-  A^{-(\delta_1+\delta_2)\sqrt{n}}\, \delta_1\delta_2\, e^{-\delta_1^2/\eps - \delta_2^2/(1{-}\eps)}
+ A^{-2\,\delta_2\sqrt{n}}\,\delta_1^2\,e^{-\delta_1^2/\eps(1{-}\eps)}
\RB .
\label{eqnA23}
\end{eqnarray}

\subsection{The probability density of the top path}

The probability that the top path passes through a vertex with
coordinates $(\lfl \eps n \rfl,\lfl \delta \sqrt{n} \rfl)$ is obtained by integrating
equation \Ref{eqnA23} for $\delta_1\in (0,\delta_2)$, and then determining
the normalising factor $N_o$.  Simplifying $F(\eps)\,F(1{-}\eps)$ gives
\begin{equation}
a_0 = F(\eps) F(1{-}\eps)
= \frac{4 \log^4A}{\pi n^3 \sqrt{\eps^3(1{-}\eps)^3}} 
\, 16^n\, e^{(n\log^2 A)/4} .
\label{eqnA24}
\end{equation}

Next, the four terms in parenthesis in equation \Ref{eqnA23} must be 
integrated for $\delta_1\in[0,\delta_2]$.  Label these terms in order by 
$[a_1,a_2,a_3,a_4]$.
Integrating and simplifying $a_1$ gives
\begin{equation}
\int_0^{\delta_2} a_1 \, d\delta_1
= \frac{\delta_2^2 (1-A^{-2\,\delta_2 \sqrt{n}})}{2\sqrt{n} \log A}\,
e^{-\delta_2^2/\eps(1{-}\eps)} .
\end{equation}
The integration of $a_2$ gives a more complicated result
\begin{equation}
\int_0^{\delta_2} a_2 \, d\delta_1
= \frac{\delta_2^2}{\sqrt{n}\,\log A}\,A^{-2\,\delta_2 \sqrt{n}}
   \,e^{-\delta_2^2/\eps(1{-}\eps)}
  - \frac{\delta_2}{n\log^2A}\, A^{-\delta_2\sqrt{n}}\,e^{-\delta_2^2/\eps}   .
\end{equation}
Integrating $a_3$ gives a similar result:
\begin{equation}
\int_0^{\delta_2} a_3 \, d\delta_1
= \frac{\delta_2^2}{\sqrt{n}\,\log A}\,A^{-2\,\delta_2 \sqrt{n}}
   \,e^{-\delta_2^2/\eps(1{-}\eps)}
  - \frac{\delta_2}{n\log^2A}\, A^{-\delta_2\sqrt{n}}\,e^{-\delta_2^2/(1{-}\eps)}   .
\end{equation}
Integrating the last term gives
\begin{equation}
\int_0^{\delta_2} a_4 \, d\delta_1
= \frac{\eps(1{-}\eps)}{4} A^{-2\,\delta_2 \sqrt{n}}
\LB\sqrt{\eps(1{-}\eps)}\,\erf\LB \delta_2/\sqrt{\eps(1{-}\eps)}\RB
-2\,\delta_2\,e^{-\delta_2^2/\eps(1{-}\eps)} \RB .
\end{equation}
These integrals must be combined with $a_0$ in equation \Ref{eqnA24}
and integrated for $\delta_2\in(0,\infty)$ to determine the normalising
constant in equation \Ref{eqnA24}.  Executing the integrals, and simplifying
gives
\begin{equation}
N_o^{-1} = \frac{\log^3A }{2\sqrt{\pi n^7}}\, 16^n\, e^{n(\log^2A)/4}.
\end{equation} 
Combining these results, putting $\delta_2=\delta$, and keeping only 
the leading terms, give the asymptotic probability for the top path 
passing through a vertex $(\lfl \eps n \rfl,\lfl \delta \sqrt{n} \rfl)$:
\begin{eqnarray}
P_n^{A,(top)}(\eps,\delta)
&\simeq \frac{4\,\delta^2}{\sqrt{\pi \eps^3(1{-}\eps)^3}} 
e^{-\delta^2/\eps(1{-}\eps)}\nonumber \\
&\hspace{1cm}
 - \frac{8\,\delta}{\sqrt{\pi n\eps^3(1{-}\eps)^3}\, \log A}
\LB e^{-\delta^2/\eps} + e^{-\delta^2/(1{-}\eps)} \RB .
\end{eqnarray}
Taking $n\to\infty$ for $A>1$ gives the exact probability of the top path passing
through vertices $(\eps,\delta)$:
\begin{equation}
\mathP_r^{A,(top)} (\eps,\delta) 
= \frac{4\, \delta^2}{\sqrt{\pi\eps^3(1{-}\eps)^3}}\, e^{-\delta^2/\eps(1{-}\eps)} .
\label{eqnA31}
\end{equation}
A contourplot of this probability density is shown in figure \ref{F8}.  

The probability density $\mathP_r^{A,(top)} (\eps,\delta)$  of the top path
in the adsorbed phase can be used to calculate the probabilities that the
top path passes through a given subset of $\RealN^2$, independent of the
trajectory of the bottom path.

As in equation \Ref{eqn9A}, $R=(x_1,x_2)\times(y_1,y_2) \subset \RealN^2$ 
is an open rectangle in the plane with closure $\o{R}$.   Put 
$S=[0,1]\times[0,\infty)$.  Define $\mathP_r^{A,(top)} (\eps,\delta) = 0$
if $(\eps,\delta)\not\in S$, and define the set function
\begin{equation}
\rho R = \rho \o{R} = \int_{R\cap S} \mathP_r^{A,(top)} (\eps,\delta)\,
 d\delta\,d\eps .
\end{equation}
Notice that $\rho$ is a set function on the semialgebra of rectangles in 
$\RealN^2$.  Following the arguments towards equation \Ref{eqn10A},
$\rho$ induces an outer measure on subsets of $\RealN^2$ and is a 
probability measure on the $\sigma$-algebra of $\lambda$-measure sets 
$E$ defined by
\begin{equation}
\rho E 
          = \int_{E\cap S} \mathP_r^{A,(top)} (\eps,\delta)\, d\lambda .
\label{eqnA33}
\end{equation}
It follows that $\rho S = \rho \RealN^2 = 1$, and that
$\frac{d\rho}{d\lambda} = \mathP_r^{A,(top)} (\eps,\delta)$ is the
Radon-Nikodym derivative of $\rho$ with respect to $\lambda$.  In this
case $\rho E$ is the probability that the top path passes through the set $E$
along its trajectory from $(0,0)$ to $(1,0)$, irrespective of the trajectory of
the bottom path.

\subsubsection{The most likely and mean trajectories of the top path in the
adsorbed phase:} 
\begin{figure}[t]
\begin{center}
\includegraphics[width=0.66\textwidth]{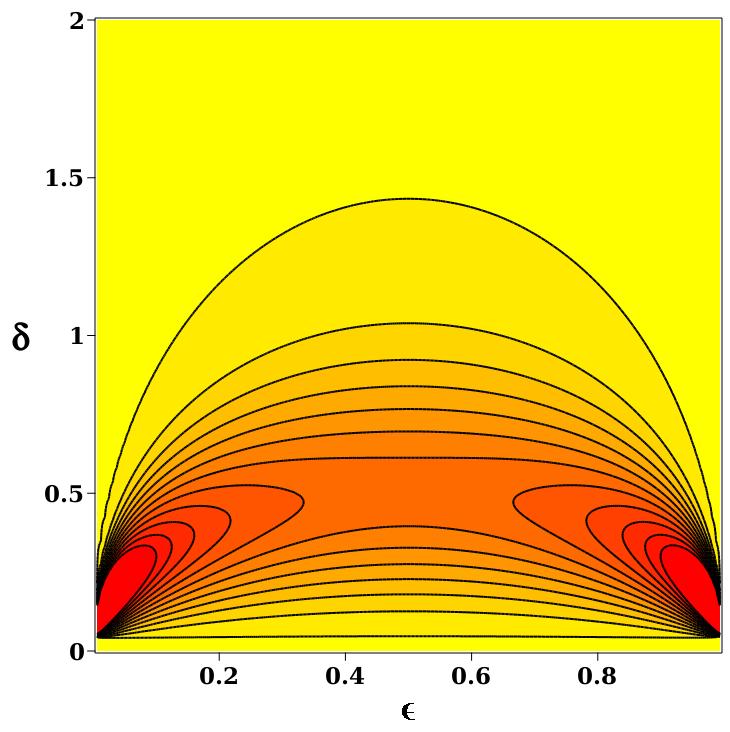}
\end{center}
\caption{Contourplot of the density $\mathP^{A,(top)}_r (\eps,\delta)$
of the top path in a staircase polygon when the bottom path is adsorbed
in the adsorbing wall (see equation \ref{eqnA31}). Lighter regions have lower densities. The plot shows low densities adjacent to the hard wall which 
increases with $\delta$ before it drops of farther from the wall.  Since the 
lower path is adsorbed on the hard wall, this density of the top path is 
the same as the density for a single Dyck path determined in 
reference \cite{JvR23}.}
\label{F8}
\end{figure}

Notice
that $\mathP_r^{A,(top)}(\eps,\delta)$ is also the probability density of a
single (desorbed) Dyck path in the scaling limit (see reference \cite{JvR23}).  
This is not unexpected, since the bottom path is adsorbed onto the 
hard wall, and the top path is screened from the wall, and does not interact 
with it.  Figure \ref{F8} shows a low density region next to the hard wall.  
The most likely trajectory of the top path here is $\delta_M = 
\sqrt{\eps(1{-}\eps)}$ \cite{JvR23}, while its mean trajectory is
$\o{\delta} =(2/\sqrt{\pi})\sqrt{\eps(1{-}\eps)} $.

\subsection{The probability density of both paths}

The asymptotic approximation to the probability density that either the top 
path, or the bottom path, passes through a given vertex is obtained by 
integrating equation \Ref{eqnA23} to $\delta_2$ in $[0,\infty)$ (or equivalently,
to $\delta_1$ in $[0,\infty)$).  As before, the normalising factor $N_o$ must be
calculated, and $a_0$ in equation \Ref{eqnA24} is independent of $\eps$,
$\delta_1$ and of $\delta_2$.

\begin{figure}[t]
\begin{center}
\includegraphics[width=0.66\textwidth]{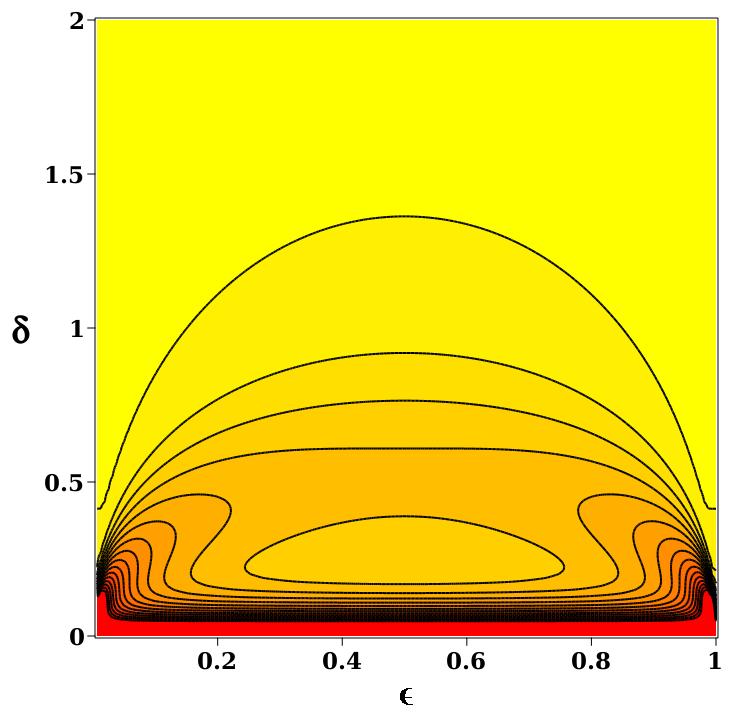}
\end{center}
\caption{Contourplot of the density $P_n^{A,(both)} (\eps,\delta)$
(equation \Ref{eqnX26}) for $a=2.5$ and $n=400$.  The bottom path is adsorbed,
while the top path is screened from the adsorbing wall by the bottom path.  
Between these paths is a low density region.}
\label{F9}
\end{figure}

Continue by integrating equation \Ref{eqnA23} for $\delta_2$ from 
$0$ to $\infty$ and then putting $\delta_1=\delta$.  This gives
\begin{eqnarray*}
& \hspace{-2cm}
N_o\,F(\eps)F(1{-}\eps)\,
\LB 
\sqrt{\pi\,\eps(1{-}\eps)}\,A^{-2\,\delta\sqrt{n}} / 4 
+\delta^2\,e^{-\delta^2/\eps(1{-}\eps)} / (2 \log(A) \sqrt{n} ) \right. \cr
& \hspace{-1.75cm}
\left.
-A^{-\delta \sqrt{n}}(\delta/4)\,e^{-\delta^2/(1{-}\eps)} 
\left(  2\,\eps - \sqrt{\pi n \eps^3}\,\log A\,e^{(n\eps \log^2 A)/4}
 \,\erfc\LB (\sqrt{\pi n\eps}\, \log A)/2 \RB
\right) \right. \cr
&  \hspace{-1.75cm}
\left.
-A^{-\delta\sqrt{n}}(\delta/4)\,e^{-\delta^2/\eps} 
\times \right. \cr
& \hspace{-1cm}
\left.
\left(  2\,(1{-}\eps) - \sqrt{\pi n (1{-}\eps)^3}\,\log A\,e^{(n(1{-}\eps) \log^2 A)/4}
 \,\erfc\LB (\sqrt{\pi n (1{-}\eps)}\, \log A)/2 \RB
\right) \RB .
\end{eqnarray*}
If $\sqrt{n}$ is large, then the complementary error functions can be
approximated using equation \Ref{eqnA3}.  This simplifies the above to
\begin{equation}
N_o\,F(\eps)F(1{-}\eps)\,
\LB \sqrt{\pi\eps(1{-}\eps)}\, A^{-2\,\delta/\sqrt{n}}
+ 2\,\delta^2 e^{-\delta^2/\eps(1{-}\eps)}/(\sqrt{n}\log A) \RB /4 .
\end{equation}
Integrating $\delta\in(0,\infty)$ and simplifying gives
\begin{equation}
N_o^{-1} = \frac{\log^3A}{\sqrt{\pi n^7}}\,16^n\, e^{n\log^2A /4}.
\end{equation}
Substituting for $N_o$ and simplifying gives the asymptotic approximation 
of the probability that either path passes through a point with coordinates 
$(\lfl \eps n \rfl, \lfl \delta \sqrt{n} \rfl)$:
\begin{equation}
{P}_n^{A,(both)}(\eps,\delta)
\simeq \sqrt{n}A^{-2\,\delta\sqrt{n}}\log A + \frac{2\,\delta^2}{
\sqrt{\pi\eps^3(1{-}\eps)^3}}
\, e^{-\delta^2/\eps(1{-}\eps)} .
\label{eqnX26}
\end{equation}
A plot of this probability is shown for $A=1.5$ and $n=400$ in figure \ref{F9}. 
The normalisation of the two terms in equation \Ref{eqnX26} is such that
\begin{equation}
\int_0^\infty \sqrt{n}A^{-2\,\delta\sqrt{n}}\log A \, d\delta= 1/2, 
\quad\hbox{and}\quad \int_0^\infty \frac{2\,\delta^2}{\sqrt{\pi\eps^3(1{-}\eps)^3}}
\, e^{-\delta^2/\eps(1{-}\eps)}\, d\delta = 1/2 .
\label{eqnX27}
\end{equation}

Proceed by constructing a probability measure on plane measure sets in
the limit as $n\to\infty$.  The second term in equation \Ref{eqnX26} is, up to
a factor of $1/2$, the same as $\mathP_r^{A,(top)}(\eps,\delta)$ in equation
\Ref{eqnA31}.  That is, it induces the measure $\rho/2$ on $\lambda$-measure
sets in $\RealN^2$ as in equation \Ref{eqnA33}.

Next, consider the first term on the right hand side in equation \Ref{eqnX26}.
Define the open rectangle $R=(x_1,x_2)\times(y_1,y_2)$ and its closure
$\o{R}$ in $\RealN^2$ and let $S=[0,1]\times[0,\infty)$.  Define the set 
function $\alpha_n$ on rectangles $R$ by
\begin{equation}
\alpha_n R = \alpha_n \o{R}
= 2 \int_{R\cap S} \sqrt{n}\,A^{-2\,\delta\sqrt{n}}\log A\, d\delta\,d\eps .
\end{equation}

Next, take the limit that $n\to\infty$ to define the set function 
$\alpha R = \lim_{n\to\infty} \alpha_n R$.  For example, 
$\alpha S = \lim_{n\to\infty}\alpha_n S = 1$ by direct integration.

Proceed by defining the infinite half-open rectangle 
$R_\delta = (\eps_1,\eps_2)\times [\delta,\infty)$ where
$\delta>0$ and $0 < \eps_1 < \eps_2 < 1$.  By excuting the integral, 
\begin{equation}
\alpha R_\delta
= \lim_{n\to\infty} \alpha_n R_\delta 
= \lim_{n\to\infty} A^{-2\,\delta\sqrt{n}}\,(\eps_2-\eps_1) = 0,
\label{eqnA39}
\end{equation}
since $A>1$ and $\delta>0$.

Moreover, if $R_0 = (\eps_1,\eps_2)\times [0,\infty)$ then integrating
directly, $\alpha R_0 = (\eps_2-\eps_1)$.


Next, consider the rectangles $T_\delta = (\eps_1,\eps_2)\times [0,\delta)$
where $\delta>0$ and $0 < \eps_1<\eps_2 < 1$.  Then
$T_\delta = R_0 \setminus R_\delta$ and it follows that if $\delta>0$, then
\begin{equation}
\alpha T_\delta = \alpha (R_0\setminus R_\delta) = \alpha R_0 - \alpha R_\delta
= (\eps_2-\eps_1) .
\end{equation}
For all other rectangles $R$ in $\RealN^2$, define
\begin{equation}
\alpha R = \alpha(R\cap S). 
\end{equation}
Then $\alpha$ is a set-function on the semi-algebra of rectangles, and
it extends to an outer measure on subsets of $\RealN^2$ and to a
measure on the $\sigma$-algebra of $\lambda$-measure sets in $\RealN^2$.

Notice that $\alpha R = 0$ on all rectangles except on those where the
intersection $R\cap [0,1]$ is an interval $[a,b]$ (or the open or 
half-open interval), in which case $\alpha R = (b-a)$.   In addition, 
$\alpha \RealN^2=1$ so that it is a probability measure and 
it induces a probability density $\mathP_r^{(\alpha)}(\eps,\delta)$ such that
\begin{equation}
\alpha E = \int_E \mathP_r^{A,(\alpha)}(\eps,\delta)\,d\lambda
= \mu(E\cap [0,1])
\label{eqnA42}
\end{equation}
where $\mathP_r^{A,(\alpha)}(\eps,\delta)=0$ if 
$(\eps,\delta)\in\RealN^2\setminus[0,1]$, and where $\mu$ is 
Lebesgue measure.  This shows that $\mathP_r^{A,(\alpha)}(\eps,\delta)$ 
has support only on the line segment $[0,1]$ in $\RealN^2$ and is 
zero everywhere else.  That is $\alpha \RealN^2 = \mu[0,1] = 1$.

Using the probability measure $\alpha$, the probability that either the
bottom or top paths pass through a $\lambda$-measure set $E$ 
in the $n\to\infty$ limit can be calculated from equation \Ref{eqnX26}:
\begin{equation}
\lim_{n\to\infty} \int_E P_n^{A,(both)}(\eps,\delta)\, d\lambda
= \lim_{n\to\infty} \int_E \sqrt{n}\,A^{-2\,\delta\sqrt{n}}\log A\, d\lambda
+ \sfrac{1}{2} \rho E
= \sfrac{1}{2} \alpha E + \sfrac{1}{2} \rho E .
\label{eqnA43}
\end{equation}
This defines a probability measure $\kappa = (\alpha + \rho)/2$ on 
$\lambda$-measure sets such that $\kappa E$ is the probability that
either the bottom, or the top path, or both, pass through $E$.
Notice that $\alpha$ and $\rho$ are mutually singular measures since
there is a decomposition $\RealN^2 = A\cup B$ with $A\cap B =
\emptyset$ such that $\alpha A = \rho B = 1$ and $\alpha B = \rho A = 0$
(for example, choose $A=[0,1]$,

\subsubsection{Most likely trajectories in the adsorbed phase:}

The most likely trajectories of both paths can be calculated using 
equation \Ref{eqnX26} in the limit $n\to\infty$ and the measures
$\alpha$ and $\rho$ in equation \Ref{eqnA43}.  This gives the
most likely trajectory of the bottom path $\delta_M=0$ and of 
the top path is $\delta_M = \sqrt{\eps(1{-}\eps)}$, for $\eps\in[0,1]$.

In the case that $n$ is finite, the most likely trajectories of the paths 
can be approximated as follows.  The derivative of equation 
\Ref{eqnX26} simplifies to
\begin{equation}
\sfrac{d}{d\delta} P_n^{A,(both)} (\eps,\delta)
\simeq \frac{4\,\delta(\eps(1{-}\eps)-\delta^2)}{\sqrt{\pi\eps^5(1{-}\eps)^5}}
\, e^{-\delta^2/\eps(1{-}\eps)} - 2\,n\, (\log A)^2 A^{-2\,\delta\sqrt{n}} .
\end{equation}
Putting the right hand side equal to zero, solve for $\delta$. This can only 
be done approximately.  Proceed by writing this in the form
\begin{equation}
\LB \frac{4\,\delta(\eps(1{-}\eps)-\delta^2)}{\sqrt{\pi\eps^5(1{-}\eps)^5}}
\, e^{-\delta^2/\eps(1{-}\eps)} \RB^{1/n}
= (2\,n\, (\log A)^2 A^{-2\,\delta\sqrt{n}})^{1/n} .
\end{equation} 
Expand the right hand side asymptotically to first order in $n$, so that
\begin{equation}
\frac{4\,\delta(\eps(1{-}\eps)-\delta^2)}{\sqrt{\pi\eps^5(1{-}\eps)^5}}
\, e^{-\delta^2/\eps(1{-}\eps)} 
\approx \LB 1 - 2\delta(\log A)/\sqrt{n} + O(1/n) \RB^n .
\label{eqnA55}
\end{equation}
Taking $n\to\infty$ on the right hand side gives (for $\delta>0$),
\begin{equation}
\lim_{n\to\infty} (1 - 2\,\delta(\log A)/\sqrt{n} + O(1/n))^n = 0
\end{equation}
and in this case the solution of equation \Ref{eqnA55} is 
$\delta_M = \sqrt{\eps(1{-}\eps)}$, as expected. 

If $n$ is finite, proceed by assuming that to leading order, the solution 
to equation \Ref{eqnA55} is $\delta^2 = K\, \eps(1{-}\eps)$, where $K$ 
may be a function of $\eps$.  Substitute and simplify to obtain 
the approximation
\begin{equation}
\frac{4\sqrt{K}(K{-}1)\,e^{-K}}{\sqrt{\pi}\eps(1{-}\eps)}
\approx \LB 1 - 2\sqrt{K\eps(1{-}\eps)} (\log A)/\sqrt{n} + O(1/n) \RB^n .
\label{eqnA57}
\end{equation}
Examination of the above shows that there are potential solutions for 
small values of $K$, and also for $K\approx 1$.  This corresponds to the
approximate most likely trajectories of the top path in figure \ref{F9} (the 
larger value of $K$), and for the ``valley'' between the top path, and the 
finite size adsorbed bottom path in figure \ref{F9}.

In order to estimate the most likely trajectory of the path, put 
$K=1+\eta^2$ in equation \Ref{eqnA57}.  Expand the left hand side 
to $O(\eta^2)$ and the factor on the right hand side to $O(1)$ to obtain
\begin{equation}
\frac{4\,\eta^2}{e\sqrt{\pi}\,\eps(1{-}\eps)}
\approx \LB 1 - 2\sqrt{\eps(1{-}\eps)}\, (\log A)/\sqrt{n}\RB^n
\end{equation}
Solving for $\eta^2$ gives
\begin{equation}
\eta^2 \approx \Sfrac{e}{4} \sqrt{\pi}\,\eps(1{-}\eps)\,
\LB 1 - 2\sqrt{\eps(1{-}\eps)} (\log A)/\sqrt{n}\RB^n
\approx \Sfrac{1}{4} \sqrt{\pi}\eps(1{-}\eps)\,
e^{1-2\sqrt{n\,\eps(1{-}\eps)}\,\log A} .
\end{equation}
Since $\delta^2 = K\,\eps(1{-}\eps)$, and $K=1+\eta^2$, this gives
the approximate most likely trajectory for finite $n$:
\begin{equation}
\delta_M = \sqrt{1+\Sfrac{1}{4} \sqrt{\pi}\eps(1{-}\eps)\,
e^{1-2\sqrt{n\,\eps(1{-}\eps)}\,\log A}}
\; \sqrt{\eps(1{-}\eps)} .
\end{equation}
Taking $n\to\infty$ gives $\delta_M = \sqrt{\eps(1{-}\eps)}$, as
expected.  If $n=400$, $A=1.5$ and $\eps=1/2$, then this
simplifies to $\delta_M \approx 0.5$, consistent with the top path
in figure \ref{F9}.  For finite values of $n$ the most likely top
trajectory is futher away from the hard wall, consistent with it
experiencing a repulsive force from the bottom path.

The ``valley'' in figure \ref{F9} is similarly determined, but instead 
putting $K= 0+\eta^2$.  Expanding the left hand side of equation 
\Ref{eqnA57} to $O(\eta)$, and the right hand side to $O(\eta)$, give
\begin{equation}
\frac{4\eta}{\sqrt{\pi}\,\eps(1{-}\eps)}
\approx 
1 - 2\sqrt{n\eps(1{-}\eps)} (\log A) \eta .
\end{equation}
Solving for $\eta$ gives
\begin{equation}
\eta = \frac{\sqrt{\pi}\,\eps(1{-}\eps)}{
4 + 2\sqrt{\pi\,n} \sqrt{\eps^3(1{-}\eps)^3}\log A} .
\end{equation}
Thus, $K=\eta^2$ and the estimate of least likely trajectory along
the ``valley'' is 
\begin{equation}
\delta_V = \frac{\sqrt{\pi}\sqrt{\eps^3(1{-}\eps)^3}}{
4 + 2\sqrt{\pi\,n} \sqrt{\eps^3(1{-}\eps)^3}\log A} .
\end{equation}
This underestimates the location of the most likely trajectory of the valley in 
figure \ref{F9}, showing that the assumption that $\eta$ is small in 
determining $K$ is in all likelihood not applicable.  However, as $n\to\infty$,
or as $\log A \to\infty$, this converges to $\delta_V=0$,
consistent with the properties of the measures $\alpha$ and $\rho$.

The mean path is obtained by integrating $\delta\,P_n^{A,(both)}
(\eps,\delta)$ (equation \Ref{eqnX26}).  This gives
\begin{equation*}
\int_0^\infty P_n^{A,(both)}(\eps,\delta)\,d\delta
= 
\int_0^\infty \delta \sqrt{n} A^{-2\,\delta\sqrt{n}}(\log A)\, d\delta
+ \int_0^\infty \frac{2\,\delta^3}{\sqrt{\pi \eps^3(1{-}\eps)^3}}
\, e^{-\delta^2/\eps(1{-}\eps)}\,d\delta .
\end{equation*}
This gives the estimated mean path for finite $n$:
\begin{equation}
\o{\delta} = \frac{1}{4\sqrt{n}\log A} + 
  \frac{\sqrt{\eps(1{-}\eps)}}{\sqrt{\pi}} .
\end{equation}
Taking $n\to\infty$ gives the mean path 
$\o{\delta} = (1/\sqrt{\pi})\sqrt{\eps(1{-}\eps)}$.

\subsection{The probability density of the bottom path}

The starting point is again equation \Ref{eqnA23} with $a_0$ defined
as in equation \Ref{eqnA24}.  The probability will be obtained by
integrating the terms in the parenthesis in equation \Ref{eqnA23}
for $\delta_2\in[\delta_1,\infty)$ and then normalising the result.
As before, only leading order terms are kept.

\begin{figure}[t]
\begin{center}
\includegraphics[width=0.48\textwidth]{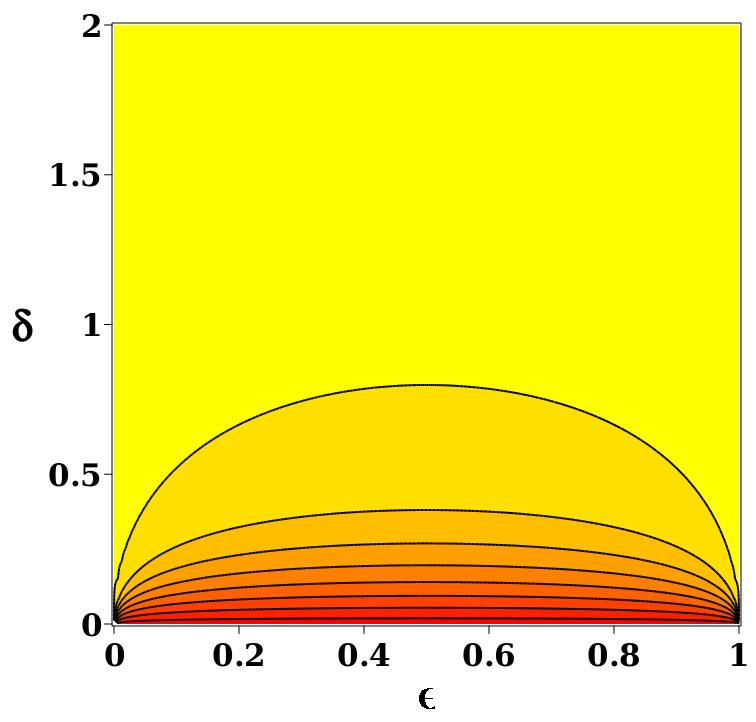}
\includegraphics[width=0.48\textwidth]{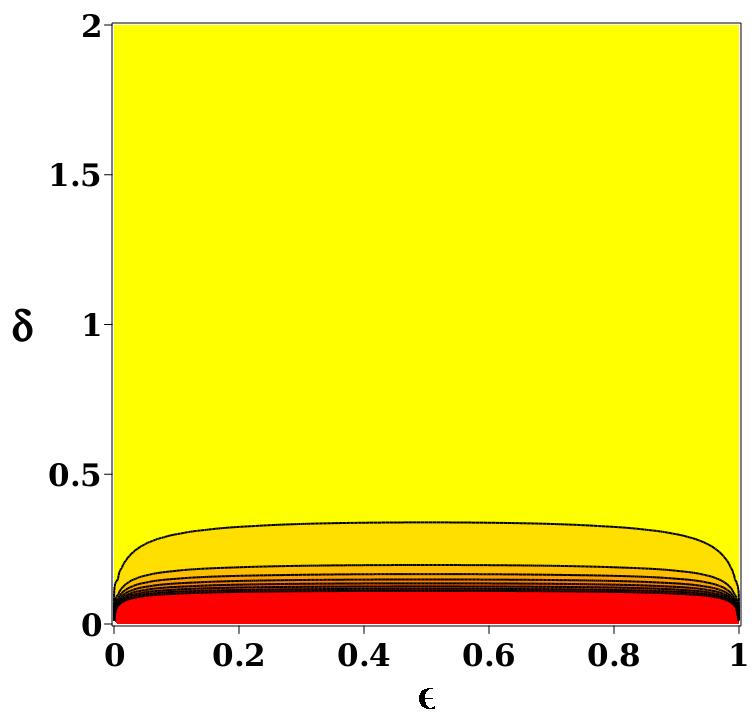}
\end{center}
\caption{Contourplot of the density $P_n^{A,(bottom)} (\eps,\delta)$
(equation \Ref{eqnA48}) for $a=2.1$ and $n=100$ (left panel) and $n=10000$
(right panel).  The bottom path is adsorbed (in the limit as $n\to\infty$).
These plots give the density of the probability that the bottom path passes
through points in the square lattice.  Darker shades correspond to higher
densities.}
\label{F10}
\end{figure}

Proceed by denoting the four terms in parenthesis in equation \Ref{eqnA23}
by $[a_1,a_2,a_3,a_4]$.  Integrating $a_1$ gives
\begin{eqnarray}
\int_{\delta_1}^\infty a_1\,d\delta_2
&= \Sfrac{1}{2}\,\delta_1\,\eps(1{-}\eps)\,A^{-2\,\delta_1\sqrt{n}}
      \,e^{-\delta_1^2/\eps(1{-}\eps)} \nonumber \\
& + \Sfrac{\sqrt{\pi}}{4}\,\sqrt{\eps^3(1{-}\eps)^3}\,A^{-2\,\delta_1\sqrt{n}}
      \,\erfc\LB \delta_1/\sqrt{\eps(1{-}\eps)}\RB .
\end{eqnarray}
The remaining three integrals give
\begin{eqnarray}
\int_{\delta_1}^\infty a_2\,d\delta_2
&= - \delta_1 \LB \Sfrac{\eps}{2} 
      + \Sfrac{\delta_1}{\sqrt{n}\,\log A}\RB A^{-2\,\delta_1\sqrt{n}}
      \,e^{-\delta_1^2/\eps(1{-}\eps)} \nonumber \\
\int_{\delta_1}^\infty a_3\,d\delta_2
&= - \delta_1 \LB \Sfrac{1{-}\eps}{2} 
      + \Sfrac{\delta_1}{\sqrt{n}\,\log A}\RB A^{-2\,\delta_1\sqrt{n}}
      \,e^{-\delta_1^2/\eps(1{-}\eps)} \\
\int_{\delta_1}^\infty a_4\,d\delta_2
&=  \Sfrac{1}{2\sqrt{n}\log A}\,\delta_1^2\,A^{-2\,\delta_1\sqrt{n}}
\, e^{-\delta_1^2/\eps(1{-}\eps)} . \nonumber
\end{eqnarray}
Adding these contributions gives the asymptotic result
\begin{equation}
P_n^{A,(bottom)} (\eps,\delta)
\simeq N_o\, \Sfrac{\sqrt{\pi \eps^3(1{-}\eps)^3}}{4}\,
A^{-2\,\delta\sqrt{n}}\, \erfc(\delta/\sqrt{\eps(1{-}\eps)})  
\end{equation}
where $N_o$ is a normalising factor.  Integrating to determine $N_o$ gives
\begin{equation}
N_o^{-1} \simeq \frac{\sqrt{\pi \eps^3(1{-}\eps)^3}}{8\sqrt{n}\log A}
 - \frac{\eps(1{-}\eps)}{8\,n\log^2 A} .
\end{equation}
Ignoring the second term and substituting gives the leading term of the probability
$P_n^{A,(bottom)}(\eps,\delta)$ that the bottom path passes through the vertex 
$(\lfl\eps n\rfl,\lfl \delta\sqrt{n}\rfl)$:
\begin{equation}
P_n^{A,(bottom)} (\eps,\delta)
\simeq 2 \sqrt{n} (\log A) A^{-2\,\delta\sqrt{n}}
\, \erfc(\delta/\sqrt{\eps(1{-}\eps)}) .
\label{eqnA48}
\end{equation}
This approximation of $P_n^{A,(bottom)}(\eps,\delta)$ is plotted in figure 
\ref{F10} for $A=1.1$ and $n=100$ (left panel) and for $n=10000$ (right panel).

A probability measure $\beta$ can now be determined similarly to 
$\alpha$ in equation \Ref{eqnA42}.  Define $S=[0,1]\times [0,\infty)$
in the $\eps\delta$-plane.  Introduce plane measure $\lambda$ and
define the $\beta$-measure of a $\lambda$-measure set $E$ by
\begin{equation}
\beta E = \lim_{n\to\infty}
\int_{E\cap S} P_n^{A,(bottom)}(\eps,\delta)\,d\delta\,d\eps
= \int_E \mathP_r^{A,(bottom)}(\eps,\delta)\,d\lambda.
\end{equation}
Then $\beta$ is a probability measure on $\RealN^2$ with density 
$\mathP_r^{A,(bottom)}(\eps,\delta)$.
 
Similar to equation \Ref{eqnA43}, for a $\lambda$-measure set E,
\begin{equation}
\beta E =
\lim_{n\to\infty} \int_{E\cap S} 2 \sqrt{n} (\log A) A^{-2\,\delta\sqrt{n}} 
\,\erfc(\delta/\sqrt{\eps(1{-}\eps)})
\, d\delta\,d\eps
\end{equation}
and it follows that $\mathP_r^{A,(bottom)}(\eps,\delta) = 0$
for $(\eps,\delta)\in \RealN^2\setminus [0,1]$.  In addition, notice
that
\begin{equation}
\lim_{n\to\infty} \int_0^\infty 2 \sqrt{n} (\log A) A^{-2\,\delta\sqrt{n}} 
\,\erfc(\delta/\eps(1{-}\eps)) \, d\delta  = 1,
\quad\forall \eps\in(0,1) . 
\end{equation}
That is, the measure $\beta$ is concentrated on the line segment $[0,1]$ 
and zero everywhere else. This shows that $\beta = \alpha$ since
$\mathP_r^{A,(\alpha)} = \mathP_r^{A,(bottom)}$. 
The most likely and mean trajectory of the bottom
path is $\delta_M=0$ and $\o{\delta}=0$ in the limit as $n\to\infty$.

\section{Probabilty densities at the critical or special point $a_c=2$}
\label{S4}

The asymptotic approximation of $R_n(h_1,h_2)$ in the case that $A=1$,
$h_1=2\lfl \delta_1 \sqrt{n} \rfl$ and 
$h_2=2\lfl \delta_2 \sqrt{n}\rfl$ is given by equation \Ref{eqn68A} in
the Appendix.  The probability that the paths pass through the points
$(2\lfl \eps n \rfl,2\lfl \delta_1 \sqrt{n} \rfl)$ and
$(2\lfl \eps n \rfl,2\lfl \delta_2 \sqrt{n} \rfl)$ is asymptotically approximated
by
\begin{equation}
P_n^{2,(sp)}(\eps,\delta_1,\delta_2) \simeq
N_o\; R_{2\,\lfl\eps n\rfl}(2\lfl \delta_1 \sqrt{n}\rfl,2\lfl \delta_2 \sqrt{n}\rfl)
\times
R_{2\,\lfl (1{-}\eps) n\rfl}(2\lfl \delta_1 \sqrt{n}\rfl,2\lfl \delta_2 \sqrt{n}\rfl)
\end{equation}
where $N_o$ is a normalising factor.  Expanding this gives to leading order
\begin{equation}
P_n^{2,(sp)}(\eps,\delta_1,\delta_2) \simeq
N_o\,\frac{\LB \delta_2^2 - \delta_1^2\RB^2}{
 \pi^2 n^5 \eps^3(1{-}\eps)^3}\,
16^{n+1}\, e^{-(\delta_1^2+\delta_2^2)/\eps(1{-}\eps)} ,
\label{eqn67A}
\end{equation}
where $N_o$ is the normalising factor which remains to be determined.

\subsection{The probability that either path pass through a given point:}

Integrating equation \Ref{eqn67A} for $\delta_2\in[0,\infty)$ and putting
$\delta_1=\delta$ gives the asymptotic probability that either the top,
or the bottom path, passes through the point $(\lfl\eps n \rfl,\lfl\delta \sqrt{n}\rfl)$:
\begin{equation}
P_n^{2,(both)}(\eps,\delta) \simeq 2\,N_o 
\frac{4\,\delta^4 - 4\,\eps(1{-}\eps)\delta^2 + 3\,\eps^2(1{-}\eps)^2}{
  n^5 \sqrt{\pi^3 \eps^5(1{-}\eps)^5} }\,
 16^n\,e^{-\delta^2/\eps(1{-}\eps)} .
\end{equation}
Normalising gives $N_o = \pi n^5\, 16^{-n-1/2}$ and the probability density
\begin{equation}
\mathP_r^{2,(both)}(\eps,\delta) = 
\frac{4\,\delta^4 - 4\,\eps(1{-}\eps)\delta^2 + 3\,\eps^2(1{-}\eps)^2}{
  2 \sqrt{\pi \eps^5(1{-}\eps)^5} }\, e^{-\delta^2/\eps(1{-}\eps)} .
\label{eqn55A}
\end{equation}
A probability measure can be defined with density similar to the way
it was defined in the free case in section \Ref{S1.2}.

\begin{figure}[t]
\begin{center}
\includegraphics[width=0.66\textwidth]{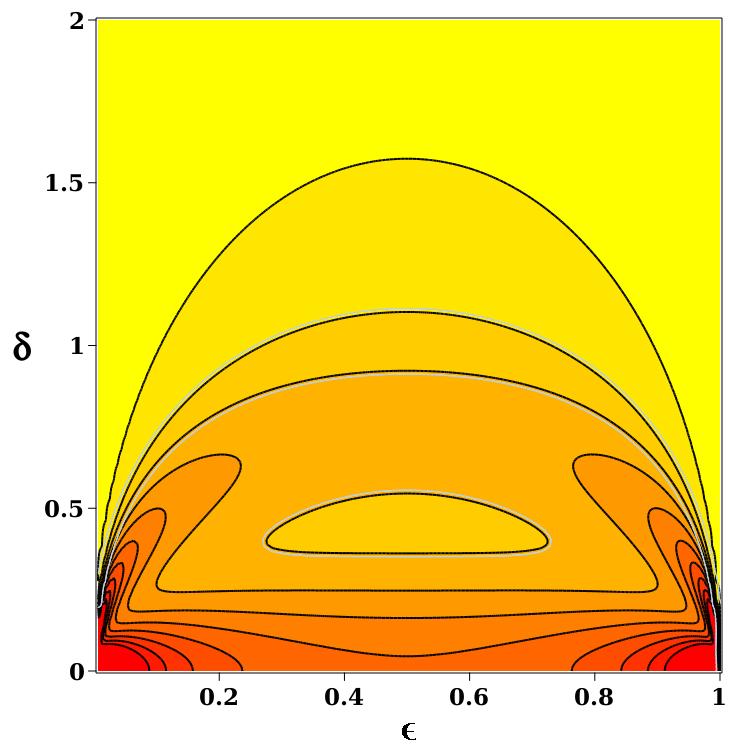}
\end{center}
\caption{Contourplot of the density $\mathP_r^{2,(both)} (\eps,\delta)$
(equation \Ref{eqn55A}) at the critical adsorption point $a_c=2$.  The 
bottom path is adsorbed, while the top path is screened from the adsorbing 
wall by the bottom path.  Between these paths is a low density region.}
\label{F12}
\end{figure}

The most likely trajectories and mean trajectories of the paths, with the 
locally least likely trajectory (the valley in figure \ref{F9}) can be 
determined as before.

Consider first the case $\mathP_2^{2,(both)}(\eps,\delta)$ in equation
\Ref{eqn55A} and figure \ref{F12}. Taking the derivative and solving
for $\delta$ gives three solutions, namely
\begin{equation}
\delta_M = 
\begin{cases}
\sqrt{\frac{3+\sqrt{2}}{2}}\, \sqrt{\eps(1{-}\eps)}, 
    & \hbox{top trajectory}; \\
\sqrt{\frac{3-\sqrt{2}}{2}}\, \sqrt{\eps(1{-}\eps)}, 
    & \hbox{valley}; \\
    0, & \hbox{bottom trajectory}.
\end{cases}
\end{equation}

The mean path can also be determined and it is
$\o{\delta} = (7/4\sqrt{\pi})\sqrt{\eps(1{-}\eps)}$.

\subsection{The probability density that the top path passes through
a given point}

The probability density that the top path passes through a given point
is obtained by integrating equation \Ref{eqn67A} for $\delta_1\in(0,\delta_2)$
and then normalising the result.  Putting $\delta_1=\delta$ gives
\begin{eqnarray}
&\hspace{-2.5cm}
P_n^{2,(top)}(\eps,\delta)
\simeq 2\,N_o \LB
\frac{4\,\delta^4-4\,\eps(1{-}\eps)\delta^2+3\,\eps^2(1{-}\eps)^2}{
n^5\sqrt{\pi^3\eps(1{-}\eps)}}\, 16^n e^{-\delta^2/\eps(1{-}\eps)}\,
\erf(\delta/\sqrt{\eps(1{-}\eps)} \right. \nonumber \\
& \hspace{3cm} + \left. \frac{2\,\delta(2\,\delta^2 -3\,\eps(1{-}\eps))}{\pi^2n^5\eps^2(1{-}\eps)^2}
 \, 16^n \,e^{-2\,\delta^2/\eps(1{-}\eps)} \RB .
\end{eqnarray}
The normalising factor is $N_o^{-1} = \pi n^5 16^{n-1/4}$.  This gives the
probability density
\begin{eqnarray}
&\hspace{-2.5cm}
\mathP_r^{2,(top)}(\eps,\delta)
= \frac{4\,\delta^4-4\,\eps(1{-}\eps)\delta^2+3\,\eps^2(1{-}\eps)^2}{
\sqrt{\pi \eps^5(1{-}\eps)^5}}\, e^{-\delta^2/\eps(1{-}\eps)}\,
\erf(\delta/\sqrt{\eps(1{-}\eps)} \nonumber \\
& \hspace{3cm}  
 +  \frac{2\,\delta(2\,\delta^2 -3\,\eps(1{-}\eps))}{\pi \eps^2(1{-}\eps)^2}
 \, e^{-2\,\delta^2/\eps(1{-}\eps)} .
\label{eqn55A}
\end{eqnarray}
This density is plotted in figure \Ref{F11}(left).

The most likely trajectory of the top path can be obtained from the 
density in equation \Ref{eqn55A}. Taking the derivative of 
$\mathP_r^{2,(top)}(\eps,\delta)$ and simplifying gives
\begin{eqnarray*}
&\hspace{-2cm} 
0 = \frac{4\,\delta^2(7\,\eps(1{-}\eps) - 2\,\delta^2)}{\pi\eps^3(1{-}\eps)^3}\,
e^{-2\,\delta^2/\eps(1{-}\eps)} \nonumber \\
&\hspace{0cm}
 + \frac{2\,\delta(4\,\delta^2(3\,\eps(1{-}\eps)-\delta^2)-7\,\eps^2(1{-}\eps)^2)
}{\sqrt{\pi\eps^7(1{-}\eps)^7}}\,
e^{-\delta^2/\eps(1{-}\eps)}\, \erf(\delta/\sqrt{\eps(1{-}\eps)}) .
\end{eqnarray*}
The zeros of the coefficients above do not coincide, but a numerical
solution can be obtained.  Assume that the trajectory is of the form
$\delta^2=K\,\eps(1{-}\eps)$ and substitute.  This gives the following
equation for $K$, independent of $\eps$:
\begin{equation}
2\sqrt{\pi K} (4\,K^2-12\,K+7)\,\erf(\sqrt{K}) + 4\,K(2\,K-7)e^{-K} = 0.
\end{equation}
There are two solutions:  $K=0$ and $K=2.2808\ldots$.  This gives
the most likely trajectory of the top path
$\delta = 1.5102\ldots \sqrt{\eps(1{-}\eps)}$.
The mean trajectory can be determined as well:
$\o{\delta} = (2\sqrt{2}/\sqrt{\pi})\sqrt{\epsilon(1{-}\epsilon)}$.

\subsection{The probability density that the bottom path passes through
a given point}

\begin{figure}[t]
\begin{center}
\includegraphics[width=0.45\textwidth]{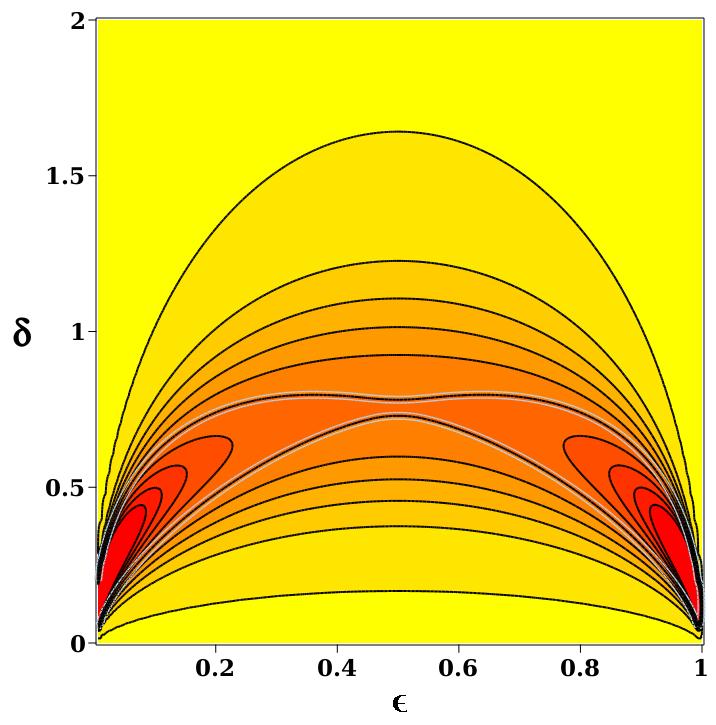}
\includegraphics[width=0.45\textwidth]{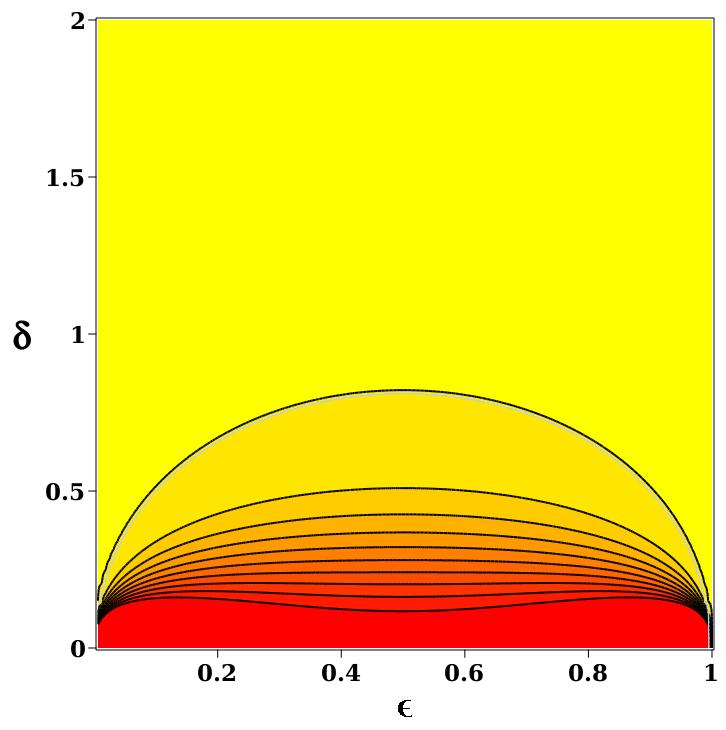}
\end{center}
\caption{Contourplots of the densities $\mathP_r^{2,(top)} (\eps,\delta)$
(left panel, equation \Ref{eqn55A}) and $\mathP_r^{2,(bottom)} (\eps,\delta)$
(right panel, equation \Ref{eqn56AX}) at the critical adsorption point $a_c=2$.
The bottom path is adsorbed the critical point (in the limit as $n\to\infty$).
Darker shades correspond to higher densities.}
\label{F11}
\end{figure}

The probability density of the bottom path is similarly obtained, and
is given by
\begin{eqnarray}
&\hspace{-2.5cm}
\mathP_r^{2,(bottom)}(\eps,\delta)
= \frac{4\,\delta^4-4\,\eps(1{-}\eps)\delta^2+3\,\eps^2(1{-}\eps)^2}{
\sqrt{\pi \eps^5(1{-}\eps)^5}}\, e^{-\delta^2/\eps(1{-}\eps)}\,
\erfc(\delta/\sqrt{\eps(1{-}\eps)} \nonumber \\
& \hspace{3cm} 
 -  \frac{2\,\delta(2\,\delta^2 -3\,\eps(1{-}\eps))}{\pi \eps^2(1{-}\eps)^2}
 \, e^{-2\,\delta^2/\eps(1{-}\eps)} .
\label{eqn56AX}
\end{eqnarray}
A contourplot is displayed in figure \ref{F11}(right). 
As before, probability measures can be defined with densities in equation
\Ref{eqn55A} and \Ref{eqn56AX}.

The most likely bottom trajectory in equation \Ref{eqn56AX} is similarly 
determined, leading to the equation
\begin{eqnarray*}
&\hspace{-2cm} 
0 = - \frac{4\,\delta^2(7\,\eps(1{-}\eps) - 2\,\delta^2)}{\pi\eps^3(1{-}\eps)^3}\,
e^{-2\,\delta^2/\eps(1{-}\eps)} + \nonumber \\
&\hspace{0cm}
 \frac{2\,\delta(4\,\delta^2(3\,\eps(1{-}\eps)-\delta^2)-7\,\eps^2(1{-}\eps)^2)
}{\sqrt{\pi\eps^7(1{-}\eps)^7}}\,
e^{-\delta^2/\eps(1{-}\eps)}\, \erfc(\delta/\sqrt{\eps(1{-}\eps)}) .
\end{eqnarray*}
Substituting $\delta_M^2=K\eps(1{-}\eps)$ and simplify.  This gives
\begin{equation}
2\sqrt{\pi K}(4\,K^2-12\,K+7)\, \erfc(\sqrt{K}) - 4\,K(2\,K-7)\,e^{-K} = 0.
\end{equation}
The only non-negative solution on the real line is $K=0$, so that
the most likely trajectory is $\delta=0$.

The mean bottom path is $\o{\delta} =
((7-4\sqrt{2})/(2\sqrt{\pi}))\,\sqrt{\eps(1{-}\eps)}$.

\begin{figure}[t]
\begin{center}
\includegraphics[width=0.66\textwidth]{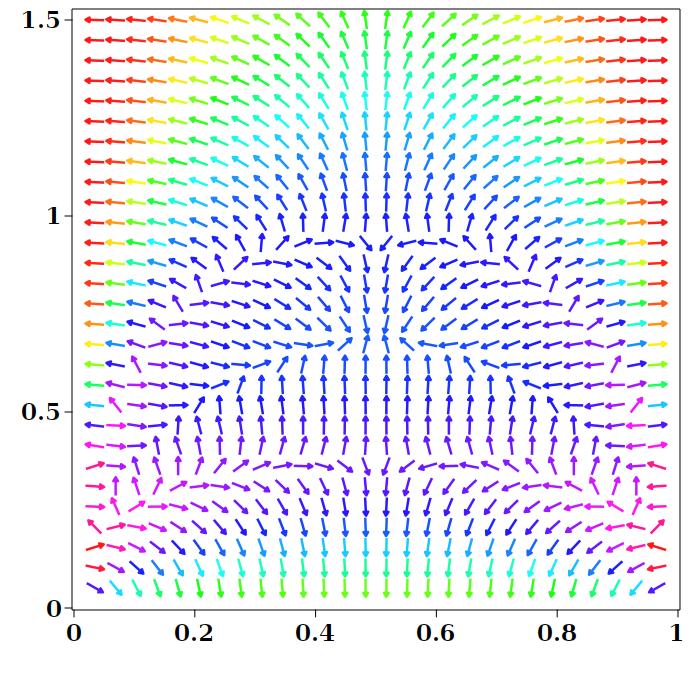}
\end{center}
\caption{A gradient plot of the probability density in the free phase
(see figure \ref{F6}).  The arrows are all of fixed length and point in 
the direction of the net force on a test particle.  The bottom 
axis is the $\epsilon$-coordinate and the vertical axis is $\delta$-coordinate,
so that the hard wall is the bottom horizontal axis. There are attractive 
forces towards to wall in its vicinity.  There are also net attractive forces 
towards the local minimum in the density near the centre of the figure. 
Towards the top of the figure the gradient shows a repulsive force 
away from the hard wall.}
\label{F13}
\end{figure}

\begin{figure}[t]
\begin{center}
\includegraphics[width=0.48\textwidth]{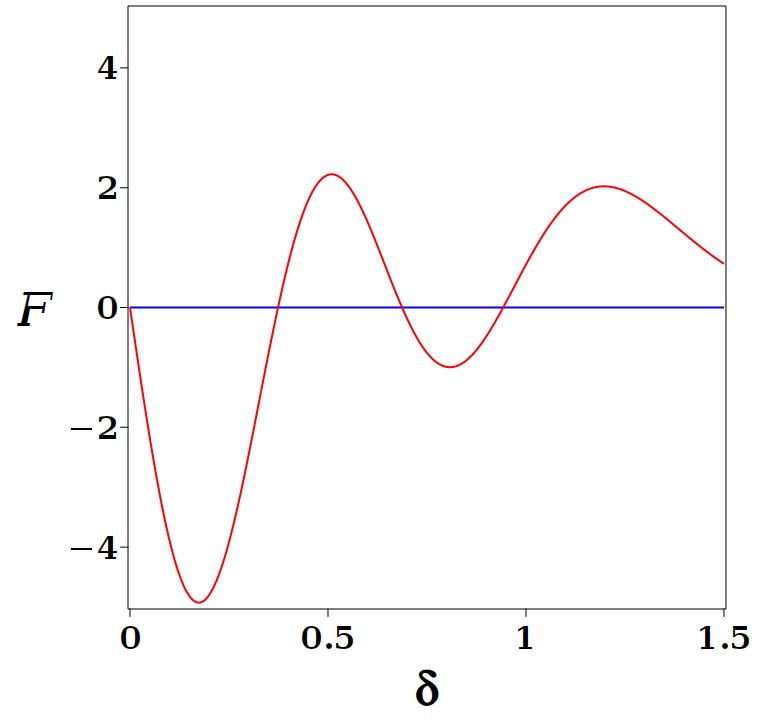}
\includegraphics[width=0.48\textwidth]{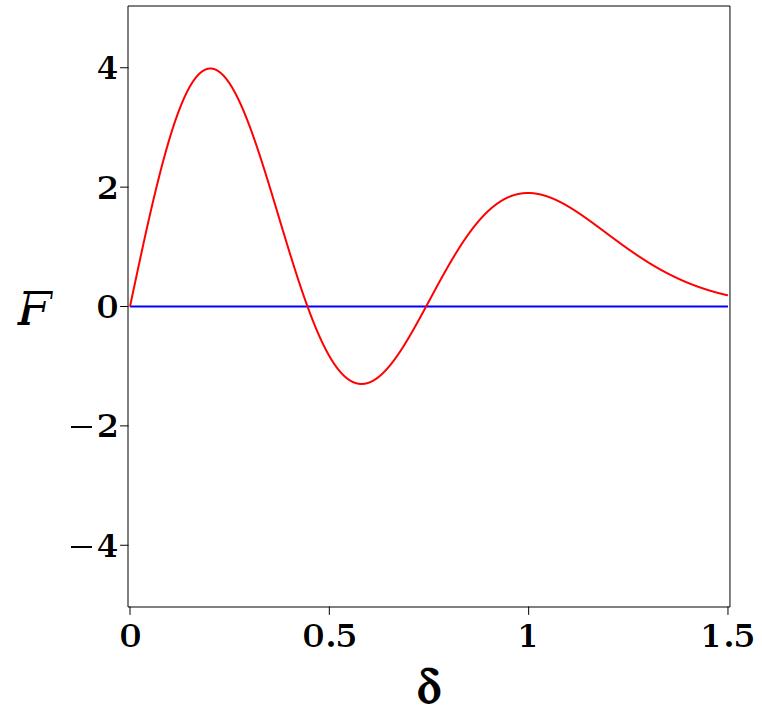}
\end{center}
\caption{(Left) The vertical component $F$ of the force in the gradient plot
in figure \ref{F13} calculated from the probability density 
$\mathP_r^{f,(both)}(\eps,\delta)$ for both paths in the free phase.
Negative values of the force are towards the hard wall, and positive
forces point away from it.  (Right)  The forces at the special point.
Here, a test particle experiences a repulsive force when placed next
to the hard wall.
}
\label{F14}
\end{figure}

\section{Discussion}
\label{S5}

Forces due to density gradients can be calculated by taking the gradient
of the probability densities.  For example, the gradient of the probability 
density in the free case for both paths in equation \Ref{eqn8A} is shown
in the gradient plot in figure \ref{F13}. This is a rescaled force-field 
experienced by a test particle placed at a vertex $(\eps,\delta)$ in the
scaling limit.  The directions of the arrows indicate the direction of
the force, and the lengths of the arrows are fixed.

Plotting the force strength in the vertical direction for $\eps=1/2$ gives
the left panel in figure \ref{F14}.  There are four points where the
forces disappear.  Numbering from origin, the zeroth and second zeros
are stable minima in the density, where a test particle will experience
restorative forces towards a stable (local) minimum. 

Calculating the forces at the special point gives the right panel in
figure \ref{F14}.  In constrast with the free case, the forces are positive
(repulsive) next to the hard wall.  There remains an interval where the
forces are attractive, and this is located between the two paths.

In this paper the most likely and mean trajectories of paths were calculated
from the exact probability distributions.  Summaries are given in 
tables \ref{T1} (most likely trajectories) and \ref{T2} (mean trajectories), for
all the cases and phases considered.  In table \ref{T1} the columns
corresponds to the most likely trajectories of the top and bottom paths,
as well as the ``valley'' separating them.  Exact values of $K$ are shown
where possible, but the numerical results in the first row were calculated
from the exact trajectories displayed in equation \Ref{eqn12Q}.

In table \ref{T2} the exact mean trajectories are shown for 
both paths, and the top and bottom paths, and in the various phases.

\renewcommand{\arraystretch}{2}
\begin{table}[h]
\caption{Most likely trajectories $\delta_M=K\sqrt{\eps({-}\eps)}$}
\centering
\scalebox{1.0}{        
\begin{tabular}{l | @{}*{3}{c}}
\br    
Model &\quad $K$ (Top path) \quad 
             & \quad $K$ (``Valley'') \quad 
                & \quad $K$ (Bottom path) \quad \cr 
\mr
Free \quad 
     & $1.8848\ldots$ & $1.3741\ldots$ & $0.7476\ldots$ \cr
Adsorbed \quad
     & $1$ & $0$ & $0$ \cr
Special point \quad
     & $\sqrt{\frac{3{+}\sqrt{2}}{2}}$ 
       & $\sqrt{\frac{3{-}\sqrt{2}}{2}}$
         & $0$ \cr 
\br
\end{tabular}
}
\label{T1}   
\end{table}

\renewcommand{\arraystretch}{2}
\begin{table}[h]
\caption{Mean trajectories $\o{\delta}=K\sqrt{\eps({-}\eps)}$}
\centering
\scalebox{1.0}{        
\begin{tabular}{l | @{}*{3}{c}}
\br  
Model &\quad $K$ (Both paths) \quad 
             & \quad $K$ (Top path) \quad 
                & \quad $K$ (Bottom path) \quad \cr 
\mr
Free \quad 
     & $\frac{5}{2\sqrt{\pi}}$ 
       & $\frac{5}{\sqrt{2\pi}}$ 
         & $\frac{5(\sqrt{2}{-}1)}{\sqrt{2\pi}}$ \cr
Adsorbed \quad
     & $\frac{1}{\sqrt{\pi}}$ 
       & $\frac{2}{\sqrt{\pi}}$
         & $0$ \cr 
Special point \quad
     & $\frac{7}{4\sqrt{\pi}}$ 
       & $\frac{2\sqrt{2}}{\sqrt{\pi}}$ 
          & $\frac{7-4\sqrt{2}}{2\sqrt{2}}$ \cr
\br
\end{tabular}
}
\label{T2}   
\end{table}

The model in this paper shows that the density of grafted staircase polygons
has low density regions.  This suggests that the monomer distribution of a 
grafted polymer network in two dimensions is uneven with low density 
regions between polymers and the hard wall, and also between 
polymers.  At sufficiently low temperatures, a test particle will 
likely localize in such a low density region until it diffuses away 
from the network due to thermal vibrations, or until the network 
degrades and releases it.  The results here show that there is 
little scope for an adsorbed polymer chain to contain a 
test particle, as its density decreases with distance from
the wall, as already seen in the model in reference \cite{JvR23}.  However,
a second (desorbed) polymer, grafted to the wall or to the first, may cause
the formation of lower density regions, as seen in figures  \ref{F9} and
\ref{F12}. More realistic simulations using Monte Carlo simulations
of self-avoiding walks or chains may give more insight into the densities
of grafted polymers, and the mechanism for capturing and stabilising 
absorbed particles in a polymer network.

\section*{Appendix:}

To determine an asymptotic approximation to $r(j,l)$ asymptotic formulae
for $n!$ and for the error function are needed:
\begin{eqnarray}
&n! \simeq n^ne^{-n}\sqrt{(2n+1/3)\,\pi}\;(1+O(1/n^2)) , 
\label{eqn56A} \\
&\erf (x) \simeq 1 - (1/(\sqrt{\pi}\,x)) \, e^{-x^2}\,(1+O(1/x^2)) . 
\label{eqnA3}
\end{eqnarray}

Proceed by substituting $n\to2n$, $j\to2j$ and $\ell\to 2\ell$ in equation \Ref{A1}.  
Replacing $m_1=k-m_2$, the sum over $m_2$ can be executed.  This gives
\begin{equation}
r_{2n}(2j,2\ell)
=\sum_{k=0}^n
\LB F_n(j,k,\ell) - G_n(j,k,\ell) \RB\, A^k
\label{eqnA4}
\end{equation} 
where $A=a-1$, and
\begin{eqnarray*}
\hspace{-2.5cm}
F_n(j,k,\ell) &= \scalebox{1.0}{
\hbox{$\dps
\frac{(2j{+}k{+}2)(n{+}1{+}2\ell(\ell{+}k{+}2){-}2j(j{+}k{+}2))
}{2(n{+}1)^2(2n{+}1)}
\binom{2n{+}2}{n{+}1{+}j{-}\ell}
\binom{2n{+}2}{n{+}3{+}j{+}k{+}\ell}$}} \\
\hspace{-2.5cm}
G_n(j,k,\ell) &= \scalebox{1.0}{
\hbox{$\dps
\frac{(2j{+}k{+}2)(n{-}1{+}2\ell(\ell{-}k){-}2j(j{+}k{+}2){-}2k)
}{2(n{+}1)^2(2n{+}1)}
\binom{2n{+}2}{n{+}2{+}j{+}\ell}
\binom{2n{+}2}{n{+}2{+}j{+}k{-}\ell} $}}
\end{eqnarray*}
The length $2n$ is the \textit{half-length} or width of the pair of walks.  
The partition function for half-length $2n$ and ending in even heights 
$h_1$ and $h_2$,  with $h_1<h_2$, is 
\begin{equation}
Z_{2n}(h_1,h_2) = r_{2n}((h_1{+}h_2)/4-1/2,(h_2{-}h_1)/4-1/2) . 
\end{equation}
For example, this shows that $Z_4(2,4) = 3+3a$.

Since the paths are ballistic in the horizontal direction, and are random 
walks in the vertical direction, the natural length scales in this model 
is $O(n)$ horizontally, and $O(\sqrt{n})$ vertically.  Thus, substitute 
$n \to \lfl \sigma n \rfl$, $k=\lfl \eta n \rfl$, 
$2(j-k)=h_1=2\,\lfl \delta_1 \sqrt{n} \rfl$ and 
$2(j+k+1)=h_2=2\,\lfl \delta_2 \sqrt{n} \rfl$.

Continue by first calculating an asymptotic approximation to $F_n(j,k,\ell)$.
Convert the binomial coefficients to factorials and use the Stirling approximation
in equation \Ref{eqn56A}.  Simplifying gives
\begin{equation}
F_n \simeq
\frac{36\, a_2 a_3 a_4 a_7 a_{20} a_0^{2\,a_0} e^{2\,a_1 + a_{18}} 
}{
\pi a_0^2 a_8 a_9^2 a_{10}^{a_{10}} a_{13}^{a_{13}} a_{15}^{a_{15}}
a_{18}^{a_{18}} \sqrt{ a_{12} a_{14} a_{17} a_{19} } \, e^{a_4+a_{11}+a_{16}}
}
\label{eqn59A}
\end{equation}
where
\begin{align*}
&a_0=2 \sigma n  {+} 3 &
&a_1={-}2 \sigma n  {-} 3 &
&a_2=36 \sigma n  {+} 57\\
&a_3=\sigma n {+} (\delta_1{+}\eta) \sqrt{n}  {+} 3 &
&a_4=\sigma n {-} \delta_2 \sqrt{n}  {+} 2 &
&a_5=(\delta_2{{-}}\delta_1) \sqrt{n} /2 {-} 1/2 \\
&a_6=(\delta_2{+}\delta_1) \sqrt{n} /2 {-} 1/2 &
&a_7=(\delta_1{+}\delta_2{+}\eta) \sqrt{n} {+} 1 &
&a_8=2 \sigma n  {+} 1 \\
&a_9=2 \sigma n  {+} 2 &
&a_{10}=\sigma n {+} \delta_2 \sqrt{n}  {+} 1 &
&a_{11}={-}\sigma n {-} \delta_2 \sqrt{n}  {-} 1 \\
&a_{12}=18\sigma  n {+} 18 \delta_2 \sqrt{n}  {+} 21 &
&a_{13}=\sigma n {-} \delta_2 \sqrt{n} {+} 2 &
&a_{14}=18\sigma  n {-} 18 \delta_2 \sqrt{n} {+} 39  \\
&a_{15}=\sigma n {+}( \eta{+}\delta_1 )\sqrt{n}  {+}  3 &
&a_{16}={-} \sigma n {-}( \eta {+}\delta_1 )\sqrt{n} {-} 3 &
&a_{17}=18\sigma  n {+} 18(\eta{+}\delta_1 )\sqrt{n}  {+} 57 \\
&a_{18}=\sigma n {-} (\eta {+} \delta_1 )\sqrt{n} &
&a_{19}=18\sigma  n {-} 18(\eta  {+} \delta_1 )\sqrt{n}  {+}  3 \\
\end{align*}
and
\begin{equation*}
a_{20}= a_6^2 + (\eta\sqrt{n} + 2)a_6 - a_5^2 + a_5\eta\sqrt{n} 
+ \eta\sqrt{n} - \sigma n/2 + 1/2 .
\end{equation*}
Take the logarithm of $F_n$ in equation \Ref{eqn59A} and expand.  
Expand the resulting expression asymptotically in $n$.  Collect the 
leading order terms to obtain
\begin{eqnarray}
&\hspace{-2.5cm}
\log F_n \simeq
4\sigma n\log 2
- (5\log n)/2
+2\log 2
- \log \pi 
-4\log \sigma
+\log (\eta+\delta_1+\delta_2) \nonumber \\
&\hspace{-1cm}
+\log (2\,\eta\delta_2+\delta_1\delta_2-\sigma)
- \eta^2/\sigma -2\,\eta\delta_1/\sigma -\delta_1^2/\sigma-\delta_2^2/\sigma .
\end{eqnarray}
Exponentiate and simplify the result.  This shows that
\begin{equation}
F_n \simeq \frac{(\eta+\delta_1+\delta_2)(2\,\eta\delta_2
            +2\,\delta_1\delta_2-\sigma)}{
\pi\sigma^4 n^{5/2}}\;
16^{\sigma n+1/2}
e^{-(\eta^2+2\,\eta\delta_1+\delta_1^2+\delta_2^2)/\sigma} .
\label{eqnA6}
\end{equation}
The sum over $k$ in equation \Ref{eqnA4} is now done by multiplying
this by $A^{\eta\sqrt{n}}$ (where $A=(a-1)$) and then integrating
for $\eta\in(0,\infty)$.

The asymptotic approximation of $G_n$ is similarly obtained, and the result is
\begin{equation}
G_n \simeq 
\frac{(\eta+\delta_1+\delta_2)(2\,\eta\delta_1+2\,\delta_1\delta_2-\sigma)}{
\pi\sigma^4 n^{5/2}}\; 
16^{\sigma n+1/2}
e^{-(\eta^2+2\,\eta\delta_2+\delta_1^2+\delta_2^2)/\sigma} .
\label{eqnA7}
\end{equation}
Notice that this is the same result as for $F_n$, but with $\delta_1$ and $\delta_2$
interchanged.

Multiplying equation \Ref{eqnA6} by $A^{\eta\sqrt{n}}$ and integrating over 
$\eta\in(0,\infty)$ gives
\begin{eqnarray*}
&\hspace{-1.5cm}
\frac{(\sigma\delta_2 n\log A 
  + \sqrt{n}(\delta_2^2{-}\sigma))\log A}{\sqrt{\pi\sigma^5 n^5}}
\nonumber \\
& \times A^{-\delta_1\sqrt{n}}\,16^{\sigma n}
e^{(\sigma^2 n (\log A)^2-4\,\delta_2^2)/4\sigma}
\LB 1+\erf\LB\frac{\sigma\sqrt{n}\log A {-} 2\,\delta_1}{2\sqrt{\sigma}} \RB \RB \nonumber \\
& \hspace{0.5cm}
+\frac{2\,\delta_2\sigma \sqrt{n} \log A + 2\,\delta_1\delta_2+2\,\delta_2^2-\sigma}{
\pi\sigma^3n^{5/2}}\,16^{\sigma n}
e^{-(\delta_1^2+\delta_2^2)/\sigma} .
\end{eqnarray*}
Expanding this asymptotically in $n$ (and using equation \Ref{eqnA3}), keeping
only leading order terms, and then simplifying produce
\begin{equation}
\sum_{k=0}^n F_n\,A^k 
\simeq \frac{2\,\delta_2 (\log A)^2}{\sqrt{\pi \sigma^3 n^3}} \,
16^{\sigma n} A^{-\delta_1\sqrt{n}}e^{\sigma n(\log A)^2/4}e^{-\delta_2^2/\sigma}.
\label{eqnA8}
\end{equation}
The similar result for $G_n$ is
\begin{equation}
\sum_{k=0}^n G_n\,A^k 
\simeq \frac{2\,\delta_1 (\log A)^2}{\sqrt{\pi \sigma^3 n^3}} \,
16^{\sigma n} A^{-\delta_2\sqrt{n}}e^{\sigma n(\log A)^2/4}e^{-\delta_1^2/\sigma}.
\label{eqnA9}
\end{equation}

\begin{figure}[t]
\begin{center}
\includegraphics[width=0.48\textwidth]{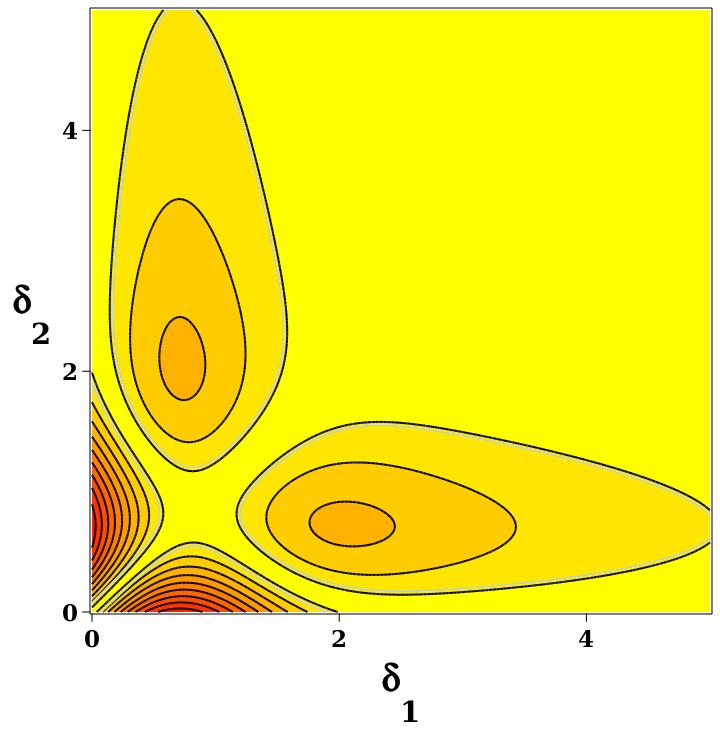}
\hfill
\includegraphics[width=0.48\textwidth]{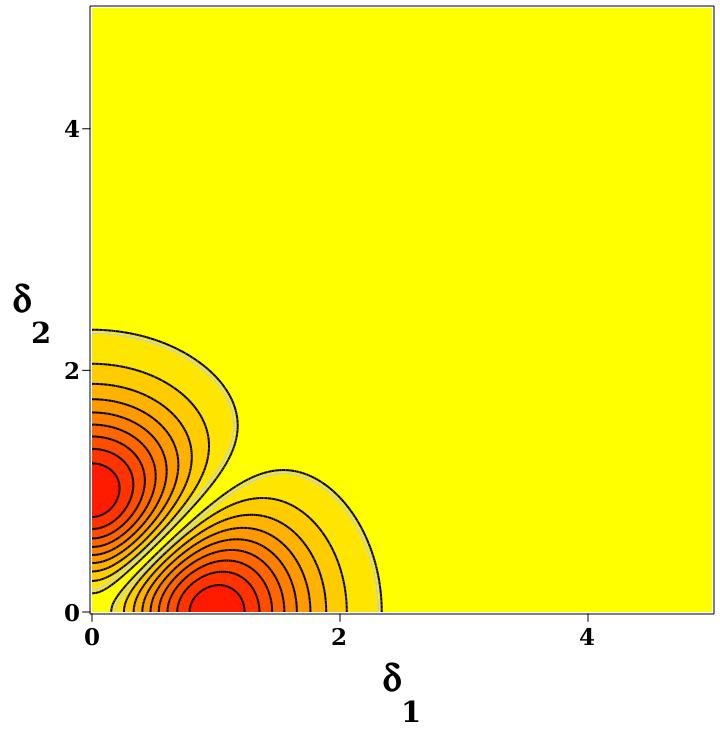}
\end{center}
\caption{Contourplots of 
$|R_{2\,\lfl \sigma n\rfl}(2\lfl \delta_1 \sqrt{n}\rfl,2\lfl \delta_2 \sqrt{n}\rfl)|$
on the $\delta_1\delta_2$-plane.  Low values are in light shades, and high values in
darker shades.
Left panel: The case that $A=1.1$ and $n=25$ while $\sigma=1.0$.  This shows 
that paths with endpoints at heights $(\delta_1,\delta_2) \approx(0,1.4)$ 
or $(\delta_1,\delta_2)\approx(1.4,0)$ dominate the counts.  There are secondary
peaks at $(\delta_1,\delta_2) \approx (0.72,2.09)$ and $(2.09,0.72)$.
Right panel:  The case that $A=1$ (at the adsorption critical point) and
$n=25$ while $\sigma=1.0$.  Paths with endpoints at heights 
$(\delta_1,\delta_2) \approx(0,1.0)$ or $(\delta_1,\delta_2)\approx(1.0,0)$ 
dominate the counts.}
\label{F15}
\end{figure}

Denoting by $R_{2n}(h_1,h_2)$ the total number of states of half-length $2n$, 
with endpoints of the paths at heights $h_1=2\lfl \delta_1 \sqrt{n}\rfl$ 
and $h_2=2\lfl \delta_2 \sqrt{n}\rfl$, then
\begin{eqnarray}
&\hspace{-2cm}
R_{2\,\lfl \sigma n\rfl}(2\lfl \delta_1 \sqrt{n}\rfl,2\lfl \delta_2 \sqrt{n}\rfl)
\simeq
\frac{2(\log A)^2}{\sqrt{\pi \sigma^3 n^3}}\,
16^{\sigma n} e^{\sigma n (\log A)^2/4}
\label{eqnA10} \\
&\hspace{3cm}
\times \LB \delta_2\,A^{-\delta_1 \sqrt{n}}\,e^{-\delta_2^2/\sigma}
- \delta_1 \,A^{-\delta_2 \sqrt{n}}\,e^{-\delta_1^2/\sigma}
\RB . \nonumber
\end{eqnarray}
In figure \ref{F15}  the absolute value
$| R_{2\,\lfl \sigma n\rfl}(2\lfl \delta_1 \sqrt{n}\rfl,2\lfl \delta_2 \sqrt{n}\rfl) |$
is plotted as a function of $(\delta_1,\delta_2)$.  The peaks in the
contour plot at $(\delta_1,\delta_2) \approx(0,1.4)$ (or 
$(\delta_1,\delta_2)\approx(1.4,0)$) corresponds to paths with $h_1=0$
and $h_2\approx 2\lfl 1.4\sqrt{n} \rfl$ and this dominates the counts
of pairs of paths.  In this figure, $n=25$, $\sigma=1$ and $A=1.1$.  There
are secondary peaks when $(\delta_1,\delta_2) \approx (0.72,2.09)$ 
and $(2.09,0.72)$.

In the case that $a=2$ consider $r_{2n}(2j,2\ell)$ in equation \Ref{eqnA4} 
with $F_n(j,k,\ell)$ and $G_n(j,k,\ell)$ as given.  Proceeding as above gives
equations \Ref{eqnA6} and \Ref{eqnA7}.  In the case that $A=a-1=1$
the approximations of the sums $F_n$ and $G_n$ over $n$ is obtained
by integrated to $\eta$.  This shows that to leading order
\begin{eqnarray}
\sum_{k=0}^n F_n 
&\simeq \frac{2\,(2\delta_1\delta_2+2\delta_2^2-\sigma)}{\pi \sigma^3 n^{5/2}} 
\, 16^{\sigma n}\, e^{-(\delta_1^2+\delta_2^2)/\sigma}, \\
\sum_{k=0}^n G_n 
&\simeq \frac{2\,(2\delta_1^2+2\delta_1\delta_2^2-\sigma)}{\pi \sigma^3 n^{5/2}} 
\, 16^{\sigma n}\, e^{-(\delta_1^2+\delta_2^2)/\sigma} .
\end{eqnarray}
In this event the approximation in equation \Ref{eqnA10} (for $A=1$) becomes
\begin{equation}
R_{2\,\lfl\sigma n\rfl}(2\lfl \delta_1 \sqrt{n}\rfl,2\lfl \delta_2 \sqrt{n}\rfl)
\simeq \frac{4\,(\delta_2^2-\delta_1^2)}{\pi \sigma^3 n^{5/2}}\,
16^{\sigma n}\,e^{-(\delta_1^2+\delta_2^2)/\sigma} .
\label{eqn68A}
\end{equation}
This is the approximation when $a=2$.  In figure \ref{F13} the absolute
value $| R_{2\,\lfl\sigma n\rfl}(2\lfl \delta_1 \sqrt{n}\rfl,2\lfl \delta_2 \sqrt{n}\rfl) |$
is plotted in the right panel.  Notice that the secondary peaks present in the
left panel have disappeared but there are peaks at $(\delta_1,\delta_2)=(0,1)$
and $(\delta_1,\delta_2)=(1,0)$.  This shows that paths with endpoints at heights
$(h_1,h_2) = (0,\lfl \sqrt{n} \rfl)$ dominates the counts.  In this figure $n=25$.

\vspace{3mm}
\noindent{\bf Data statement:}
No numerical data were generated in this study.

\vspace{3mm}
\noindent{\bf Acknowledgements:} 
EJJvR is grateful for financial support from NSERC (Canada) in the form of
a Discovery Grant RGPIN-2019-06303  , and to York University for funds 
supporting this research.

\vspace{2cm}
\noindent{\bf References}
\bibliographystyle{plain}
\bibliography{trajectory}

\end{document}